\documentclass{vgtc}                

\ifpdf
  \pdfoutput=1\relax                   
  \usepackage{graphicx}                
  \DeclareGraphicsExtensions{.pdf,.png,.jpg,.jpeg} 
\else
  \usepackage{graphicx}                
  \DeclareGraphicsExtensions{.eps}     
\fi%

\graphicspath{{figures/}{pictures/}{images/}{./}} 
\makeatletter
\def\input@path{{tables/}{./}}
\makeatother
\usepackage{microtype}                 
\PassOptionsToPackage{warn}{textcomp}  
\usepackage{textcomp}                  
\usepackage{mathptmx}                  
\usepackage{times}                     
\usepackage{cite}                      
\usepackage{tabu}                      
\usepackage{booktabs}                  
\usepackage{mathtools}
\usepackage{amsmath}
\usepackage{amssymb}
\usepackage{enumitem}
\usepackage{algorithmic}
\usepackage{adjustbox}
\usepackage{algorithm}
\usepackage{float}
\usepackage[table, dvipsnames]{xcolor}
\usepackage{tikz}
\usetikzlibrary{arrows.meta}
\usetikzlibrary{positioning}
\usetikzlibrary{calc}
\usepackage{wrapfig}

\onlineid{1000}

\vgtccategory{Research}
\vgtcpapertype{Algorithm/Technique}

\newcommand{\julien}[1]{\textcolor{black}{#1}}
\newcommand{\julesEdit}[1]{\textcolor{black}{#1}}

\title{Fast Approximation of Persistence Diagrams with Guarantees}

\author{Jules Vidal and Julien Tierny
\thanks{E-mails: \{jules.vidal, julien.tierny\}@sorbonne-universite.fr}
\\
CNRS -- Sorbonne Universit\'e
\vspace{-1.75ex}
}

\shortauthortitle{Vidal \MakeLowercase{\textit{et al.}}: Fast Approximation of Persistence Diagrams}

\abstract{
  This paper presents an algorithm for the efficient approximation of the 
  saddle-extremum persistence diagram of a scalar field. 
  Vidal et al. introduced recently a fast algorithm for such an approximation 
(by interrupting a progressive computation framework  \cite{vidal21}). However,
no theoretical guarantee was provided regarding
its approximation quality.
In this work, we revisit the progressive framework of Vidal et al. 
\cite{vidal21} and we introduce in contrast
a novel approximation algorithm, with a user 
controlled approximation error, specifically, on the Bottleneck distance to the 
exact solution.
Our approach is based on a 
hierarchical representation of the input data,
and relies on local simplifications of the scalar field to accelerate the 
computation,
while maintaining a controlled bound on the output error.
   The locality of our approach enables further speedups thanks to
   shared memory 
   parallelism. Experiments conducted on real life datasets show that for a
mild error tolerance (5\% relative Bottleneck distance), our 
approach improves runtime performance by \julien{18\%} on average (and up to
48\% on large, noisy datasets)  in
comparison to standard, exact, publicly available implementations. 
In addition to the strong 
guarantees on its approximation error, we show that our algorithm also provides
in practice
outputs which are on average \julien{5} times 
more accurate (in terms of the $L_2$-Wasserstein distance) than a naive 
approximation baseline.
We illustrate the utility of our approach for interactive data exploration and 
we document visualization strategies for conveying the uncertainty related to 
our approximations.
   
}

\teaser{
  \centering
\includegraphics[width=\linewidth]{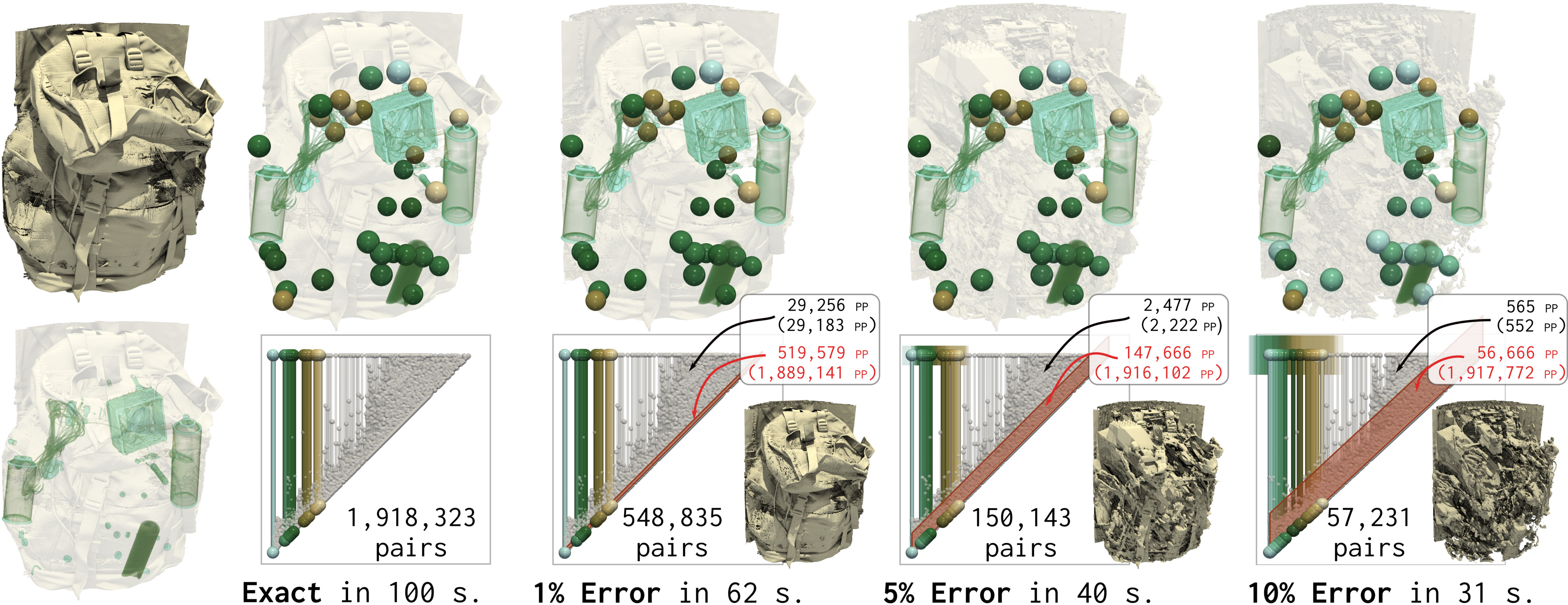}
  \mycaption{Approximations of persistence diagrams for a CT scan of a
    backpack, with different Bottleneck approximation errors.
    High persistence features correspond to high density
    objects present in the bag
    (leftmost, bottom). 
    In this example, our controlled approximation reduces computation time 
    by 70\% 
for an error tolerance
of 
10\%.
Our algorithm also provides
an approximation of the
scalar field that is precise around persistent
features (top, volume rendering of the approximated fields) and
deteriorated
elsewhere (bottom, isocontours capturing
the cloth of the bag).
For 
each 
approximation, the thirty most persistent maxima are represented by spheres (top 
views) 
which
correctly capture the features of the data. 
The uncertainty 
resulting from our approximation is 
visualized in the diagram with
\emph{(i)} colored squares, which bound the correct location of 
\emph{certain} features (which are guaranteed to be present in the exact 
result). These  are located outside a \emph{(ii)} red band (in the 
vicinity of the diagonal), which denotes features which \emph{might} not exist
in the
exact diagram.
Our approximations 
precisely capture
 the high persistence features of the data (out of the red
band, black numbers) and collect significantly less noisy features (red
numbers, red band) than the exact result (numbers in parentheses).
}
  \label{fig:teaser}
  \vspace{1ex}
}

\vgtcinsertpkg

\DeclareMathAlphabet{\mathcal}{OMS}{cmsy}{m}{n}

\begin{document}

\renewcommand{\sectionautorefname}{Sec.}
\renewcommand{\subsectionautorefname}{Sec.}
\renewcommand{\figureautorefname}{Fig.}
\renewcommand{\equationautorefname}{Eq.}
\newcommand{\algorithmautorefname}{Alg.}
\renewcommand{\tableautorefname}{Tab.}
\newcommand{\mycaption}[1]{\vspace{-3ex}\caption{#1\vspace{-2ex}}}
\newcommand{\myequation}[1]{\vspace{-2ex}#1\vspace{-1ex}}

\newcommand{\jules}[1]{\textcolor{black}{#1}}
\newcommand{\TODO}[1]{\textcolor{black}{#1}}

\firstsection{Introduction}

\maketitle

\newcommand{\domain}{\mathcal{M}}
\newcommand{\range}{\mathbb{R}}
\newcommand{\sublevelset}[1]{#1^{-1}_{-\infty}}
\newcommand{\Star}{St}
\newcommand{\Link}{Lk}
\newcommand{\simplex}{\sigma}
\newcommand{\face}{\tau}
\newcommand{\lowerlink}{\Link^{-}}
\newcommand{\upperlink}{\Link^{+}}
\newcommand{\Index}{\mathcal{I}}
\newcommand{\offset}{o}
\newcommand{\Natural}{\mathbb{N}}
\newcommand{\criticalSet}{\mathcal{C}}
\newcommand{\diagram}{\mathcal{D}}
\newcommand{\pointMetric}[1]{d_#1}
\newcommand{\wasserstein}[1]{W_#1}
\newcommand{\projection}{\Delta}
\newcommand{\hierarchy}{\mathcal{H}}
\newcommand{\decimation}{D}
\newcommand{\xDimD}{L_x^\decimation}
\newcommand{\yDimD}{L_y^\decimation}
\newcommand{\zDimD}{L_z^\decimation}
\newcommand{\xDim}{L_x}
\newcommand{\yDim}{L_y}
\newcommand{\zDim}{L_z}
\newcommand{\Grid}{\mathcal{G}}
\newcommand{\GridD}{\mathcal{G}^\decimation}
\newcommand{\x}{\phantom{x}}
\newcommand{\Mod}{\;\mathrm{mod}\;}
\newcommand{\NN}{\mathbb{N}}
\newcommand{\forwardIntegralLine}{\mathcal{L}^+}
\newcommand{\backwardIntegralLine}{\mathcal{L}^-}
\newcommand{\triangulationOp}{\phi}
\newcommand{\decimationOp}{\Pi}
\newcommand{\fold}[1]{\widehat{#1}}
\newcommand{\foldedSet}{\mathcal{F}}
\newcommand{\monotonyOffset}{\mathbb{M}}
\newcommand{\offsets}{\mathbb{O}}

\tikzset{stairwayStyleLineWidth/.style={line width=0.04em}}
\tikzset{stairwayStyleRound/.style={line join=round,line cap=round,stairwayStyleLineWidth}}
\tikzset{stairwayStyleSharp/.style={stairwayStyleLineWidth}}
\tikzset{stairwayStyle/.style={stairwayStyleRound}}
\newcommand{\stairwayuphollow}{\mathbin{
\tikz[baseline=(stairwayanchor.base)]{
    \node (stairwayanchor) {\quad}; 
        \draw[stairwayStyle] 
        ($(stairwayanchor.south west) + (0.0em,0.4em)$)
        -- ++(0,0.20em)
        -- ++(0.20em,0) -- ++(0,0.20em)
        -- ++(0.20em,0) -- ++(0,0.20em)
        -- ++(0.20em,0) 
        -- ++(0,-0.60em)
        -- cycle
        ;
}}}

\newcommand{\stairfield}{f^{\scriptsize{\stairwayuphollow}}}

\label{sec_intro}
Modern datasets, acquired or simulated, are continuously gaining in geometrical 
complexity, thanks to the ever-increasing accuracy of acquisition devices or 
computing power of high performance systems. 
This geometrical complexity makes interactive exploration and analysis 
difficult, which challenges the interpretation of large datasets by end 
users. 
This motivates the definition of expressive data abstractions, capable of 
capturing the main features present in large datasets into concise 
representations, which 
visually convey the most important information to the users.

In that context, Topological Data Analysis (TDA) \cite{edelsbrunner09} forms a 
family of generic, robust, and efficient techniques whose utility has been 
demonstrated in a number of visualization tasks \cite{heine16} for revealing 
the implicit structural patterns present in complex datasets. 
Examples of popular application fields include 
turbulent combustion \cite{laney_vis06, bremer_tvcg11, gyulassy_ev14},
 material sciences \cite{gyulassy_vis07, gyulassy_vis15, favelier16},
 nuclear energy \cite{beiNuclear16},
fluid dynamics \cite{kasten_tvcg11}, 
bioimaging \cite{carr04, topoAngler},
quantum chemistry \cite{chemistry_vis14, harshChemistry, Malgorzata19} or 
astrophysics \cite{sousbie11, shivashankar2016felix}.
%
%
These applications
rely on established 
topological data abstractions, such as 
contour trees 
\cite{boyell63, carr00, smirnov17, gueunet_tpds19}, 
Reeb graphs 
\cite{reeb1946points, biasotti08, pascucci07}
or Morse-Smale complexes \cite{Defl15, gyulassy_vis08, 
robins_pami11, gyulassy_vis18}. 
In particular, the Persistence diagram \cite{edelsbrunner02} is a 
concise data representation, which visually summarizes the 
population of features of interest present in large datasets, as a function of 
a measure of 
importance called \emph{topological persistence}. Its conciseness 
made it increasingly popular in 
data visualization and analysis,
where it quickly provides visual hints regarding the 
number and salience of the structural features present in large datasets.

While topological methods usually have an acceptable time complexity,  
the construction of the above data representations can still be time consuming 
in practice for large, real-life datasets.
Thus, when they are integrated into larger analysis pipelines,
TDA algorithms can become a serious time bottleneck. Several research 
directions are possible to improve the time performance of TDA algorithms. 
A natural avenue for performance improvement consists in revisiting these 
algorithms for a parallel computation \cite{MaadasamyDN12, ShivashankarN12,
GueunetFJ16, CarrWSA16, gyulassy_vis18, gueunet_tpds19, 
GueunetFJT19}, which is particularly relevant for high-performance hardware. 
Another direction consists in considering \emph{degraded} computations, which 
trade accuracy for speed. This direction is particularly appealing for 
persistence diagrams, as they can contain, for noisy datasets, many features 
with low persistence which would gain in being only \emph{approximated}, as 
they often have in practice only little relevance (important features typically 
have a high persistence, see \autoref{fig:teaser}).

In this work, we introduce an algorithm for the fast approximation of 
extremum persistence diagrams of large scalar datasets, with a user-controlled 
error bound. In particular, we re-visit the progressive computation framework 
of Vidal et al. \cite{vidal21} and derive a multiresolution strategy for
accelerating the computation, based on a controlled degradation of the input 
data (hence resulting in a controlled approximation of the result). Extensive 
experiments demonstrate the time performance gain provided by our algorithm (up 
to 48\% on large, noisy datasets) for 
mild
error tolerance (e.g. 5\%), as well as its superior accuracy in
comparison to a naive
approximation baseline. We also document several visualization strategies for 
our approximated topological features, to easily enable the visual
identification of
\emph{certain} features (which will be present in the exact diagram) along with 
visual clues of their positional uncertainty in the diagram.

\subsection{Related Work}
\label{sec_relatedWork}

The literature related to our work can be classified into two main
groups:
(i) topological methods, and (ii) 
multiresolution methods.

\noindent
\textbf{\emph{(i)} Topological methods } have gained a growing interest from 
the visualization
community for the last two decades \cite{heine16}. Specifically, our work is closely related
to the construction of topological data abstractions on scalar data, that provide a generic and robust
description of the features in the data.
When considering discretized data as piecewise-linear (PL) scalar fields
defined on PL-manifolds, many concepts
from the original Morse theory \cite{milnor63} can be adapted to the computational setting.
As such, combinatorial algorithms have been developed to efficiently compute
topological abstractions in the discrete setting. A local, combinatorial
characterization of the critical points of a PL scalar field was introduced by
Banchoff \cite{banchoff70}. Critical points often correspond to features of
interest in the data, but can be numerous in the presence of noise. Topological
persistence \cite{edelsbrunner09, edelsbrunner02}
addresses
this issue by
introducing an importance measure on critical points. It can be visualized in
the Persistence diagram (\autoref{sec_persistence_diagrams}), a visual barcode
that can be generally computed using matrix reduction operations
\cite{edelsbrunner09, edelsbrunner02}. The Persistence diagram has been shown
to be stable \cite{CohenSteinerEH05,Cohen-Steiner2010} under well-established
metrics such as the Bottleneck \cite{CohenSteinerEH05} and
Wasserstein \cite{Turner2014} distances. However, some
applications
call for more discriminative topological abstractions. Merge trees and contour
trees, for instance, track the merge and split events of connected components
of sub-level sets and level sets that happen at the critical points.
Combinatorial algorithms were first designed to compute these trees in low
dimension \cite{boyell63}. An efficient algorithm
was later introduced by Carr et al. \cite{carr00} for the computation of the
contour tree in all dimensions, with optimal time-complexity. A lot of work
also focused on the parallel computation of these structures \cite{smirnov17,
gueunet_ldav17, gueunet_tpds19, PascucciC03, MaadasamyDN12, MorozovW14,
CarrWSA16}. The Reeb graph \cite{reeb1946points} is the
generalization of contour trees to domains that are not simply connected, which
motivates its use for shape analysis applications \cite{biasotti08}. In order
to track the connected components of
level sets on such domains, original
algorithms resorted to slicing strategies \cite{BiasottiFS00}. Later
work optimized the slicing approach to restrict it to iso-contours at critical
points \cite{PataneSF08, DoraiswamyN12}. Algorithms with optimal time
complexity were introduced for the 2D case \cite{ColeMcLaughlinEHNP03} and in
arbitrary dimension \cite{Parsa12}. Other work focused on its parallel
computation \cite{GueunetFJT19} or presented efficient algorithms for its
computation in specific settings \cite{pascucci07, tierny_vis09}. An other
widely used topological abstraction is the Morse-Smale complex \cite{Defl15},
which encodes the relations between critical points in terms of unique integral
lines of the gradient field. These integral lines segment the domain into
cells in which
the gradient integrates to
identical
extremities. Algorithms
for its computation were designed for PL manifolds \cite{EdelsbrunnerHNP03}, or later \cite{gyulassy_vis08, robins_pami11} for the
 Discrete Morse Theory \cite{forman98}. Parallel methods have
been documented \cite{ShivashankarN12, gyulassy_vis18}.
For the case of multivariate scalar data, recent work
\cite{CarrD14,tierny_vis16} investigated efficient algorithms for the
Reeb space \cite{EdelsbrunnerHP08} computation, a generalization of the
Reeb graph.

\noindent
\textbf{\emph{(ii)} Multiresolution hierarchical representations} have been 
largely used by the visualization
community to compute and store data representations at different levels of
details \cite{weiss_sgp09, weiss_vis09}. They have been notably applied to the efficient and 
robust extraction of isosurfaces \cite{PascucciB00,
GregorskiDLPJ02,GerstnerP00} and isocontours \cite{LewinerVLM04}.
Such representations have been
also documented for
the contour tree \cite{pascucci_mr04}, the Reeb graph \cite{hilaga:sig:2001}
and the Morse-Smale complex
\cite{BremerEHP03, gunther2012, IuricichF17}. These representations allow to adaptively
simplify these topological abstractions in a \textit{fine-to-coarse} manner. As such,
only the output data structures are processed hierarchically, while the input
data are processed with classical algorithms on
the finest level of the hierarchy.

More closely related to our work, Vidal et al. \cite{vidal21} recently introduced novel ideas
for the \textit{progressive} computation of
critical points and saddle-extremum persistence diagrams.
In their approach, the input data is processed in a \textit{coarse-to-fine} manner
with the help of a multiresolution hierarchy on the model of the classical \textit{red} subdivision
\cite{freudenthal42,bank83,loop87,zhang95,bey95} of triangulated regular grids \cite{kuhn60, bey98}.
The output of the approach is progressively and efficiently updated from one
level of the hierarchy to 
the next.
In practice, their algorithms are \textit{interruptible}, which means that the
processing of the hierarchy can be stopped before reaching the finest level and
still deliver an exploitable result in a lower amount of computation time. Although the authors
demonstrated the
quality of the interrupted outputs
experimentally,
there are
no theoretical
guarantees on their approximation quality (specifically, no error bound).
In contrast,
we present in this work a method for the approximate computation
of
a persistence diagram, with strong guarantees on the
approximation
error.
%

\subsection{Contributions}
\label{sec_contributions}

This paper makes the following new contributions:

\begin{enumerate}[leftmargin=1em]
\vspace{-1.1ex}
\setlength\itemsep{0em}
\item \emph{A fast approximation algorithm for extremum persistence diagrams, 
with controlled error:}
We introduce a fast
algorithm for the approximated computation of persistence diagrams of
extremum/saddle pairs.
Our approach provides strong, user-controlled guarantees on the Bottleneck 
distance between the approximated diagram
and the exact result. By construction, our approximation method processes a 
smaller number of topological events than
exact methods, which reduces its run time. We observe an 18\% reduction of the
run time on average, for a mild tolerance of 5\% on the relative Bottleneck
error. Moreover, our approach provides much more accurate approximations 
(in terms of the $L_2$ Wasserstein distance) than a naive 
approximation baseline.
\vspace{-.25ex}





\vspace{-.25ex}
\item \emph{An approximated visualization of topological features with 
uncertainty assessment:} We describe several strategies for the visualization 
of our topological approximations. As a by-product of our algorithm, we provide 
an approximated scalar field, for visualization purposes, which exactly matches
our approximated diagrams and in which the approximated critical points can be 
reliably embedded. We augment the approximated diagrams with visual glyphs 
assessing the uncertainty of the approximation.
These glyphs enable the visual identification of 
\emph{certain} features (which will be present in the exact diagram) along with 
visual clues on their positional uncertainty in the diagram.

\vspace{-.25ex}
\item \emph{A reference implementation (additional material):}
Finally, we provide a reference C++ implementation of our algorithms that can 
be used to replicate our results as well as for future benchmarks.

\end{enumerate}


\section{Preliminaries}
\label{sec_preliminaries}
This section presents the theoretical background of our work. It contains 
definitions adapted from the Topology ToolKit \cite{ttk17}. 
We refer the 
reader to textbooks \cite{edelsbrunner09} for an introduction to 
topology.

\subsection{Input Data}
The input data is a piecewise linear (PL) scalar field $f:\domain\rightarrow\range$
defined on a PL $d$-manifold $\domain$, with $d\leq3$ in our applications.
The field $f$ is characterized by its values at the vertices of $\domain$, as scalar values are linearly 
interpolated on the simplices of $\domain$ of higher dimension.
$f$ is enforced to be injective on the vertices of $\domain$,
using a symbolic perturbation inspired by Simulation of Simplicity\cite{edelsbrunner90}.

\subsection{Critical Points}
\label{sec_critical_points}

We define a \textit{sub-level} set of $f$ as the pre-image of $(-\infty,w)$ by $f$, noted $\sublevelset{f}(w)=\left\{p\in\domain~|~f(p)<w\right\}$.
When continuously increasing $w$, the number of topological features of $\sublevelset{f}(w)$ (its connected components, independant cycles and voids) changes
only at specific vertices, called the \textit{critical points} of $f$ \cite{milnor63}.

A local characterization of critical points, based on their \textit{link}, was introduced by Banchoff\cite{banchoff70}.
For two simplices $\tau, \sigma \in \domain$, we note $\tau<\sigma$ if $\tau$ is a \textit{face} of $\sigma$, \textit{i.e.} if $\tau$ is defined by a non-empty, strict subset of the vertices of $\sigma$.
We call \textit{star} of a vertex $v\in\domain$ the set of its co-faces, noted $\Star(v)=\{\sigma\in\domain~|~v<\sigma\}$.
The \textit{link} of $v$, noted $\Link(v)=\{\tau\in\domain ~|~ \tau<\sigma~,~\sigma\in\Star(v)~,~\tau\cap v=\emptyset \}$, 
is defined as the set of all the faces $\tau$ of the simplices of $\Star(v)$ that do not intersect $v$.
Intuitively, the link of $v$ is the \textit{boundary} of a small simplicial neighborhood $\Star(v)$ of $v$.
The \textit{lower link} $\lowerlink(v)$ is the set of simplices of $\Link(v)$ that are \textit{lower} than $v$ in terms of $f$ values: 
$\lowerlink(v) = \{ \simplex \in \Link(v) ~ | ~ \forall v' \in \sigma, ~ f(v') < f(v)\}$.
Symmetrically, the \textit{upper link} of $v$ is given by 
$\upperlink(v) = \{ \simplex \in \Link(v) ~ | ~ \forall v' \in \sigma, ~ f(v') > f(v)\}$.
If both $\lowerlink(v)$ and $\upperlink(v)$ are non-empty and simply connected,
$v$ is a \textit{regular} point. In any other case, $v$ is a critical point, of
\textit{criticality}
or \textit{critical index} $\Index(v)$: 0 for a local minimum
($\lowerlink(v)=\emptyset$), $d$ for a local
maximum ($\upperlink(v)=\emptyset$), or $i$ for a $i$-saddle ($0<i<d$).

\subsection{Persistence Diagrams}
\label{sec_persistence_diagrams}
\begin{figure}
        \centering
        \includegraphics[width=0.95\linewidth]{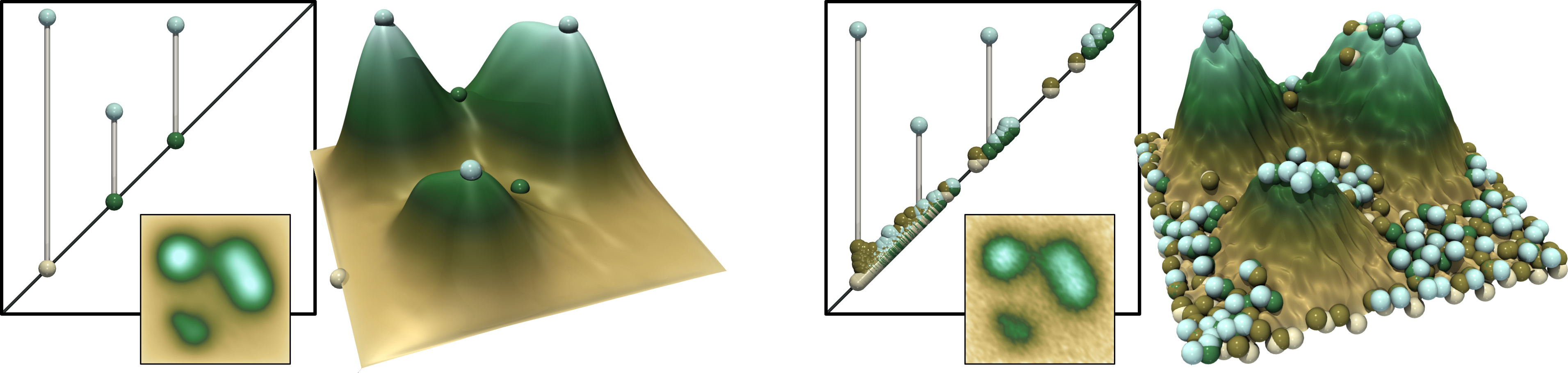}
        \mycaption{Persistence diagrams of a clean (left) and noisy (right)
 2D scalar 
 field.
Critical points are represented by spheres
 (light brown: minima, cyan: maxima, others: saddles). The persistence
 diagram highlights the three main features of the data (high persistence
 pairs). On the right, the small pairs near the diagonal indicate the presence of 
 noisy features with low persistence.
}
\label{fig_persistence_diagram}
\end{figure}

As an isovalue $w$ continuously increases, the topology of $\sublevelset{f}(w)$ evolves.
For a regular point $v$, the sub-level sets enter $\Star(v)$ through $\lowerlink(v)$
and exit through $\upperlink(v)$ without any local change in the topology, as $\lowerlink(v)$ and $\upperlink(v)$
both have exactly one connected component. Otherwise, if $v$ is critical, a topological feature (a
connected component, a cycle or a void) of $\sublevelset{f}(w)$ is either
created or destroyed at the value $f(v)$.
For instance, the connected components of $\sublevelset{f}(w)$ are created at
the local minima of $f$, said to be their \textit{birth} points.
When two topological features meet at a critical point, we use the Elder rule \cite{edelsbrunner09}
to state that the younger one (the one created last, in terms of $f$ values) \textit{dies}
at the benefit of the oldest. In particular, the connected components of $\sublevelset{f}(w)$ die at
1-saddles. 
Each topological feature of $\sublevelset{f}(w)$ is thus characterized by a unique pair
$(c_0,c_1)$ of critical points, corresponding to its \textit{birth} and \textit{death}, 
with $f(c_0)<f(c_1)$ and $\Index(c_0)=\Index(c_1)-1$. The
\textit{topological persistence}\cite{edelsbrunner02} of this pair,
noted $p(c_0,c_1)=f(c_1)-f(c_0)$, denotes the lifetime of the corresponding
feature in $\sublevelset{f}(w)$ in terms of its scalar range.
The persistence of
the connected components in $\sublevelset{f}(w)$ is encoded by minimum/1-saddle pairs.
In 3D,
$(d-1)$-saddle/maximum pairs characterize the persistence of the voids of
$\sublevelset{f}(w)$ while 1-saddle/2-saddle pairs encode the lifetime of its
independent cycles.
In practice, the features of interest in the data are often characterized by the extrema of the field $f$.
Thus we will only focus on extremum/saddle pairs in the following.


The topological persistence of each critical pair gives a measure of importance
on the corresponding extremum of the field, that has been shown to be reliable
to distinguish between noise and important features. The \textit{Persistence
diagram}\cite{edelsbrunner09} of $f$, noted $\diagram(f)$, is a visual
representation of the ensemble of features in the data, where each persistence
pair $(c_0,c_1)$
is embedded as a point in the 2D plane, at coordinates $(f(c_0),f(c_1))$.
The persistence of each pair can thus be read in the diagram as the
height of the point to the diagonal.
Consequently, each topological feature of
$\sublevelset{f}(w)$ can be visualized in the diagram as a bar (\autoref{fig_persistence_diagram}), whose height indicates its importance in the data.
Large bars corresponding to high persistence features stand out visually, while low persistence pairs, likely to be associated 
with noisy features, are represented by small bars in the vicinity of the diagonal.
The persistence diagram is a concise visual depiction of
the repartition of features in the data
and has been shown to be a stable\cite{CohenSteinerEH05} and useful tool for
data summarization tasks.
As seen in \autoref{fig_persistence_diagram}, it encodes the number, ranges and salience of
features of interest, and gives hints about the level of noise in the data. 

The \textit{Wasserstein distance} is a well-established metric
between
persistence diagrams.
Originally from the field of Transportation theory,
it is defined as the cost of an optimal assignment between
 two diagrams $\diagram(f)$ and $\diagram(g)$, where the cost of matching a
 point $a=(x_a,y_a)\in\diagram(f)$ to a point $b=(x_b,y_b)\in\diagram(b)$
 is given by a pointwise distance $d_q$ in the 2D birth/death space of the diagrams:
\begin{equation}
\pointMetric{q}(a,b)=\left(|x_b-x_a|^q + |y_b-y_a|^q\right)^{1/q} = \|a-b\|_q
\label{eq_pointWise_metric}
\end{equation}
By convention, $\pointMetric{q}(a,b)=0$ if both $a$ and $b$ are on the diagonal ($x_a=y_a$ and $x_b=y_b$).
The $L_q$-Wasserstein distance between $\diagram(f)$ and $\diagram(g)$ is
then given by:
\begin{equation}
    \wasserstein{q}\big(\diagram(f), \diagram(g)\big) = 
\min_{\phi
\in \Phi} \left(\sum_{a \in \diagram(f)} 
\pointMetric{q}\big(a,\phi(a)\big)^q\right)^{1/q}
\end{equation}
where $\Phi$ is the set of all possible maps $\phi$ mapping each
point
$a \in \diagram(f)$ to 
a point
$b 
\in \diagram(g)$
or to 
its projection onto the diagonal.
When $q$ converges to infinity, the Wasserstein distance converges to the
\textit{Bottleneck} distance, another practical metric that measures the worst
mismatch between
 $\diagram(f)$ and $\diagram(g)$:
 
\begin{equation}
    \wasserstein{\infty}\big(\diagram(f), \diagram(g)\big) = 
\min_{\phi
\in \Phi} \left(\max_{a \in \diagram(f)} 
\|a-\phi(a)\|_\infty\right)
\end{equation}
The persistence diagram is \textit{stable} under these two metrics, which intuitively means
that a small variation in the scalar field entails a small difference in the resulting distances. In particular, 
the Bottleneck distance between two diagrams is bounded by the $L_\infty$
distance between the corresponding scalar fields\cite{CohenSteinerEH05}:
\begin{equation}
    \wasserstein{\infty}\big(\diagram(f),\diagram(g)\big)\leq \|f-g\|_\infty
\end{equation}

\begin{figure}[t]
      \centering
        \includegraphics[width=\linewidth]{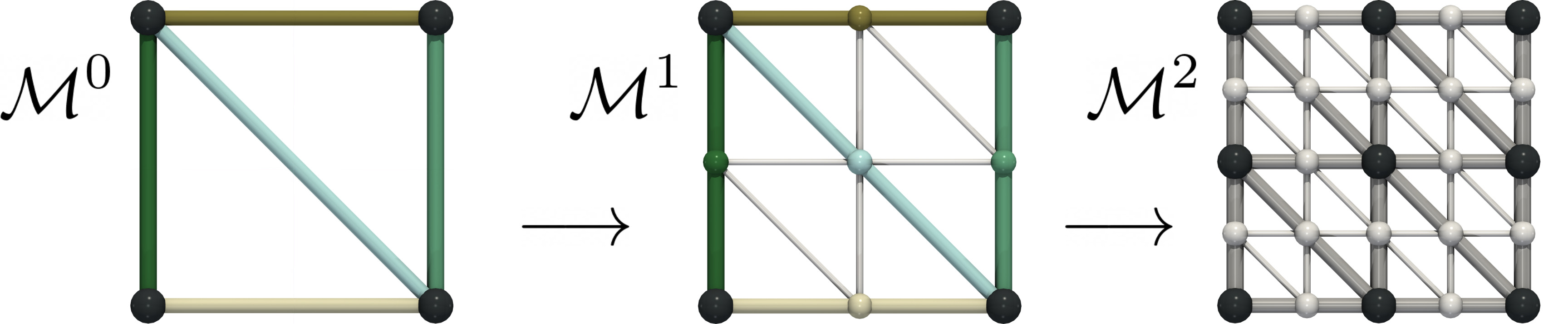}
        \mycaption{Edge-nested triangulation hierarchy
        of
        a regular
        grid.
        New vertices are only inserted
        on existing edges, one per edge, splitting each old edge in two.
        Old and new edges are shown  (right, $\domain^2$) in grey and white
respectively; old and new vertices are shown in black and white
respectively.
      New edges only connect new vertices.
    }
\label{fig_edge_nested}
\end{figure}
\begin{figure}[t]
    \centering
    \includegraphics[width=\linewidth]{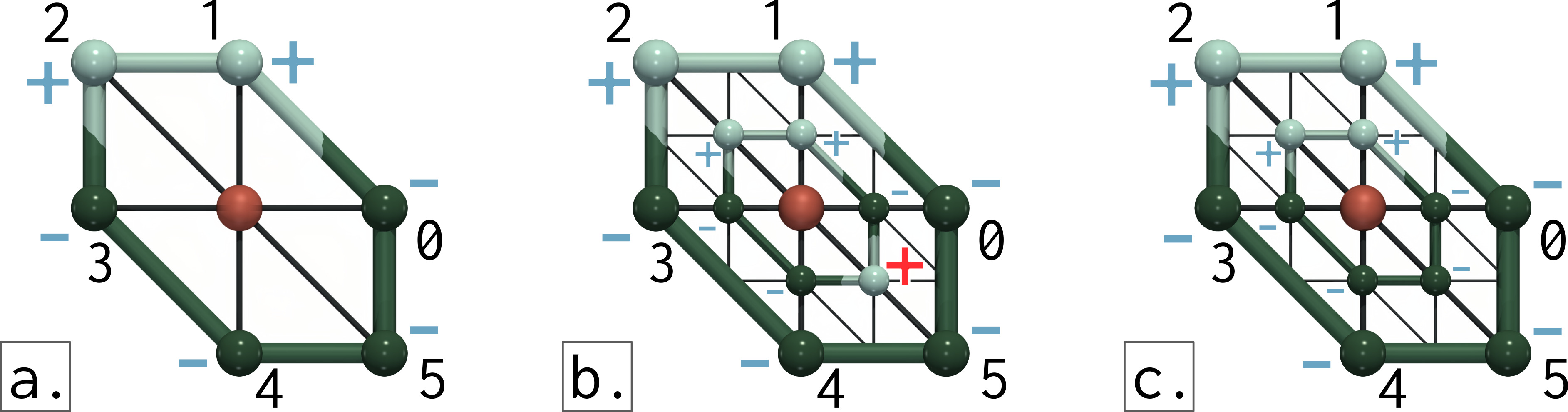}
    \mycaption {
        Evolution of the upper link (light green) and
        the lower link (dark green) of a vertex $v$ (red sphere) when
progressing
down the hierarchy. The topological structure of the link at
        level $i$ (a) remains unchanged at level $i+1$, thanks to the
        edge-nested property of the hierarchy.
In the example
        (b), neighbor 5
        is \emph{non-monotonic}: it switches its polarity (from blue minus to
red plus), which modifies the
        criticality of $v$ (saddle point in $\domain^{i+1}$).
        If the link polarity of $v$ is \emph{not} changed (c), $v$ is called a
\textit{Topologically Invariant} vertex
        and it maintains its criticality.
}
    \label{fig_topological_invariants}
\end{figure}

\subsection{Edge-Nested Hierarchical Data Representation}
\label{sec_prelis_hierarchy}

To design progressive algorithms for topological data analysis,
Vidal et al.\cite{vidal21} used a progressive representation of the input data. 
This representation is
based on an \textit{edge-nested} multiresolution hierarchy of the input domain $\domain$.
The hierarchy $\hierarchy = \{\domain^0, \domain^1, \dots, \domain^h\}$ is a
growing
sequence of PL manifolds,
such
that all the vertices present in one level $\domain^i$ of $\hierarchy$
are present in all the following levels
$\domain^j$, $j>i$.
This naturally yields a hierarchy of PL scalar fields $\{f^0,\hdots,f^h\}$
such as each vertex maintains the same scalar value from its level of insertion
$i$ forward:
$\forall v \in\domain^i,~\forall j~|~i\leq j\leq h,~ f^j(v)=f^i(v)=f(v)$.
The \textit{edge-nested} property states that \emph{new} vertices inserted at a
level $i$ of the hierarchy are exclusively inserted
on existing edges (connecting \emph{old} vertices) at level $i-1$, one per
edge, and that new edges only connect new vertices (see
\autoref{fig_edge_nested}).
We refer the reader to Vidal et al. \cite{vidal21} for a detailed
formalization.
Edge-nested hierarchies can  be easily generated from regular grids.
Given a
$d$-dimensional regular grid $\Grid^0$,
a valid PL $d$-manifold $\domain^h$ can be obtained from $\Grid^0$ by
considering its Kuhn triangulation\cite{kuhn60} (\autoref{fig_edge_nested}).
A sequence $\{\Grid^0, \Grid^1, \dots,\Grid^h\}$ of decimated grids can be recursively defined from $\Grid^0$,
where $\Grid^i$ is obtained for $i>0$ by sub-sampling only vertices with even coordinates in $\Grid^{i-1}$.
The edge-nested triangulation hierarchy
$\hierarchy = \{\domain^0, \domain^1, \dots, \domain^h\}$
is then defined
by considering the
triangulation
$\domain^i$ of each level $\Grid^{h-i}$. Further implementations details  can
be found in Vidal et al. \cite{vidal21}.


\subsection{Topological Invariant Vertices}
\label{sec_prelis_topological_invariants}
The advantage of using an edge-nested triangulation hierarchy to process the
data resides in the fact that topological information
can be computed on the vertices of $\domain^i$ and remain valid throughout the rest of the hierarchy.
In particular, as discussed by Vidal et al. \cite{vidal21}, when progressing
from a level of the hierarchy to the next, the
topological structure of the link is invariant for all old vertices: their
number of neighbor vertices remains the same,
and the adjacency relations between neighbors are preserved.
This important property (\autoref{fig_topological_invariants})
enables the efficient identification of vertices that do no change
criticality when progressing to the next hierarchical level.

For each vertex $v$ of $\domain^i$, its \textit{link polarity}
(blue $+/-$ glyphs in \autoref{fig_topological_invariants})
indicates for each one of its neighbors $n\in\Link(v)^i$, whether
$n\in\upperlink(v)^i$ (positive polarity) or $n\in\lowerlink(v)^i$ (negative
polarity).
As the structure of $\Link(v)^i$ is invariant, the criticality of $v$ is characterized by the polarity of its link. In particular, 
if the link polarity of $v$ stays unchanged when progressing a level down the
hierarchy, the criticality of $v$ remains the same and
does not need to
be recomputed.
Given an edge $(v_0,v_1)$ of $\domain^i$ that gets subdivided into two old edges $(v_0,v_n)$ and $(v_n,v_1)$ along the new vertex $v_n$ of $\domain^{i+1}$,
$v_n$ is said to be \textit{monotonic} if $f(v_n)\in(f(v_0),f(v_1))$, assuming that $f(v_0)<f(v_1)$. Otherwise, $v_n$ is called a
\textit{non-monotonic} vertex, and the polarity of the link of $v_0$ or $v_1$ is necessarily changed, as shown in \autoref{fig_topological_invariants}b. 
On the opposite, an old vertex keeps the same link polarity (and
thus the same criticality) if all its new neighbors are \textit{monotonic} (\autoref{fig_topological_invariants}c). It 
is then called a \textit{Topologically Invariant old vertex}.
Potentially, depending on the regularity of the data, a large part of the old
vertices at level $i$ can be topologically invariant.
Similarly, the criticality of a part of the \textit{new} vertices can be deduced from the hierarchy with the identification of 
non-monotonic vertices. It can be shown\cite{vidal21} that if a new vertex $v$ of $\domain^i$ is monotonic and if all its \textit{new} neighbors
are also monotonic, then $v$ is necessarily a \textit{regular} point. It is
called a \textit{Topologically Invariant new vertex}.

It has been shown\cite{vidal21} that Topologically Invariant vertices (TIs)
represent a large part of the data in practice (around 70\% on average on a
diverse sample of real-life data sets). 
Thus, the
edge-nested hierarchical representation of the data
enables a highly efficient detection of 
these TI vertices.
It proved itself useful to 
carry out topological data analysis tasks 
such as the extraction of critical points or the computation of a persistence diagram, as
significant shortcuts can be made in the computation for TI vertices.

\section{Overview}
\label{sec_preliminaries}

\begin{figure*}
  \centering
  \includegraphics[width=\linewidth]{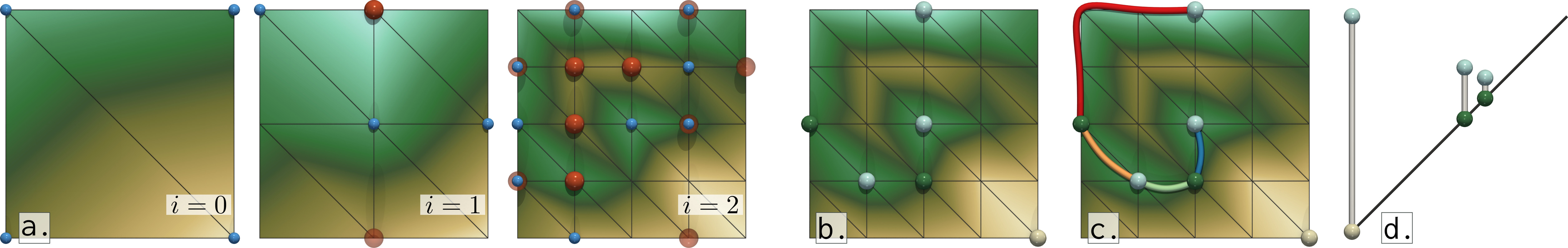}
  \mycaption{Overview of our approach. First, the traversal of the hierarchy
(a)
      enables the efficient detection of the vertices
      (blue spheres) that are \textbf{not}
topologically
      invariant (TI), and for which the criticality must be computed.
      Non-monotonic vertices (red spheres) can be \textit{folded} (transparent
      red spheres)
  during the traversal, \textit{i.e.} reinterpolated to artificially
  increase the number of TI vertices.
  Second (b), the criticality of all non-TI vertices detected in the first step
is
computed,
to identify the critical points of the approximated field
  (spheres, cyan: maxima, green: saddles,
beige:
  minimum).
  Third, the saddle
points are used to seed integral lines
  ending at extrema (c), from which the persistence diagram is deduced (d).}
  \label{fig_overview}
\end{figure*}

\autoref{fig_overview} provides an overview of our approach, which revisits
the progressive framework of Vidal et al. \cite{vidal21}, 
to derive a 
fast approximation algorithm with strong guarantees. First, we exploit their 
multiresolution hierarchy of the input data (\autoref{sec_prelis_hierarchy}) to 
quickly update, down to the finest hierarchy level,
the polarity of each vertex (a local information used to identify 
critical points). This step is described in \autoref{sec_hierarchy_processing}. 
During the hierarchy traversal, 
in contrast to their original approach, we artificially increase the number of 
\emph{topologically invariant vertices} 
(\autoref{sec_prelis_topological_invariants}) in order to significantly speedup 
the computation, through a procedure called \emph{vertex folding}, which 
artificially  degrades the input data. This step is described in 
\autoref{sec_vertex_folding}. The data degradation induced by the vertex 
folding procedure is precisely controlled in the process, to provide strong 
guarantees on the approximation error of the output. This is described in 
\autoref{sec_error_control}. Finally, we describe in 
\autoref{sec_monotonyOffsets} how to handle degenerate configurations such as 
flat plateaus.
Overall, our new approach involves three major differences to the progressive 
framework of Vidal et al. \cite{vidal21}, which are detailed in the rest of 
this section:
\begin{enumerate}[leftmargin=.3cm]
 \item Our new approximation algorithm is \emph{not} progressive: it does not 
generate a sequence of progressively refined outputs. Instead, our traversal of 
the multiresolution hierarchy only updates a minimal amount of information 
(the vertex polarity). The criticality of each vertex is only evaluated
\emph{after} the hierarchy traversal is finished
(while the criticality
is updated at each hierarchy level in \cite{vidal21}).
  \item Our approximation approach is based on a multiresolution degradation of 
the input data, which accelerates the overall computation, while maintaining a 
controlled output error.
  \item Overall, in contrast to the progressively interrupted results of 
Vidal et al. \cite{vidal21}, our approximated output is provided with strong 
guarantees on its approximation error (in terms of relative Bottleneck 
distance) and with a significantly improved practical accuracy  (in terms of 
the $L_2$ Wasserstein distance).
\end{enumerate}

\section{Topology Approximation}
\label{sec_uc1}

This section presents our novel approach for the controlled approximation of an
extremum/saddle persistence diagram.
In the following, we focus on the case of minimum/1-saddle pairs
($(d-1)$-saddle/maximum pairs being treated symmetrically).
Our method is based on \textit{folding} operations for non-monotonic vertices.
As the hierarchy is processed, 
non-monotonic vertices are inserted at each level.
When a non-monotonic vertex is inserted
in the hierarchy on a new edge, we can 
purposely decide
to reinterpolate this vertex to enforce its monotony, and hence accelerate
the computation of its criticality.
The resulting error is
the difference between the real scalar value at this vertex and the
interpolated value, which bounds the Bottleneck error on the diagram
estimation, as detailed next. 

\subsection{Hierarchy Processing}
\label{sec_hierarchy_processing}

This section presents our computation strategy in the case where the
approximation error $\epsilon$ is set to $0$ (i.e. exact computations).
Given an input edge-nested triangulation hierarchy $\hierarchy = \{\domain^0,
\domain^1, \dots, \domain^h\}$, persistence diagrams are evaluated in three
steps.
%
First, the hierarchy is completely traversed,
from the coarsest
level $\domain^0$ to the finest $\domain^h$, to efficiently detect topologically
invariant vertices, for which 
the criticality can be efficiently estimated in a second step, at the last
level of the hierarchy only ($\domain^h$).
Third, the persistence diagram is computed
from the saddle points identified at the second step.

\noindent
\textbf{1) Hierarchy traversal:} 
In order to identify topologically invariant vertices, we compute
the link polarity (\autoref{sec_prelis_topological_invariants})
of the vertices for each level of the hierarchy. The link polarity of a vertex $v$ is encoded in 
our setting as an array of bits, one per neighbor $n$ of $v$, denoting whether
$n$ is higher or lower than $v$.
The
size of the link polarity is the same for each level of $\hierarchy$
(\autoref{sec_prelis_topological_invariants}): at
most 6 bits in 2D and 14 bits in 3D.
At each level $i$, the polarity of new vertices is
initialized, while the polarity of old vertices is updated to account for the
insertion of non-monotonic vertices. The detection of non-monotonic vertices 
enables the fast identification of regular points:
\textit{new} topologically
invariant vertices are known to be regular points of $f^i$, and \textit{old}
topologically invariant vertices keep the same criticality at
level $i$ than at level $i-1$ (\autoref{sec_prelis_topological_invariants}).
We leverage these informations to avoid the explicit computation of the criticality for some of the vertices.
As a new topologically invariant vertex $v$ is identified in $\domain^i$, it is
flagged as a regular vertex. If its link polarity is not changed in the
remaining levels (e.g.
$v$ is topologically invariant through the rest of the hierarchy), $v$ is
guaranteed to be a
regular vertex of $\domain^h$ and no further computation will be needed for it.

\noindent
\textbf{2) Critical points:}
At  $\domain^h$,
we compute explicitly
the criticality of the vertices which are not yet guaranteed to be regular
 (as described above).
The criticality of a vertex $v$ is computed by enumerating the
connected
components of $\upperlink(v)$ and $\lowerlink(v)$
(\autoref{sec_critical_points}).

\noindent
\textbf{3) Persistence diagram:} 
The minimum/1-saddle persistence diagram is deduced from the critical points.
In particular, for each saddle point $s$, a backward integral line is launched
from each connected
component of $\lowerlink(s)$, if there are more than one.
These integral lines end at a local minimum $m$.
For each integral line, we back-propagate the
vertex index of $m$ back to $s$. This way, we build for each saddle $s$
the list $\{m_0, m_1, \hdots, m_{j}\}$ of found local minima, yielding at least a minimum per
connected component of sub-level sets merging at $s$. A merge event can thus be
represented by a triplet $(s,m_i,m_j)$.
 All found triplets are then sorted in ascending
values of $s$, and the merge events are processed in this order with
a UnionFind (UF) data structure \cite{cormen}. A persistence pair is created
when two different
components are merged. We refer to Vidal et al. \cite{vidal21} for more details,
as
this last step of our method is identical to theirs.

\begin{figure*}
        \centering
        \includegraphics[height=0.18\linewidth]{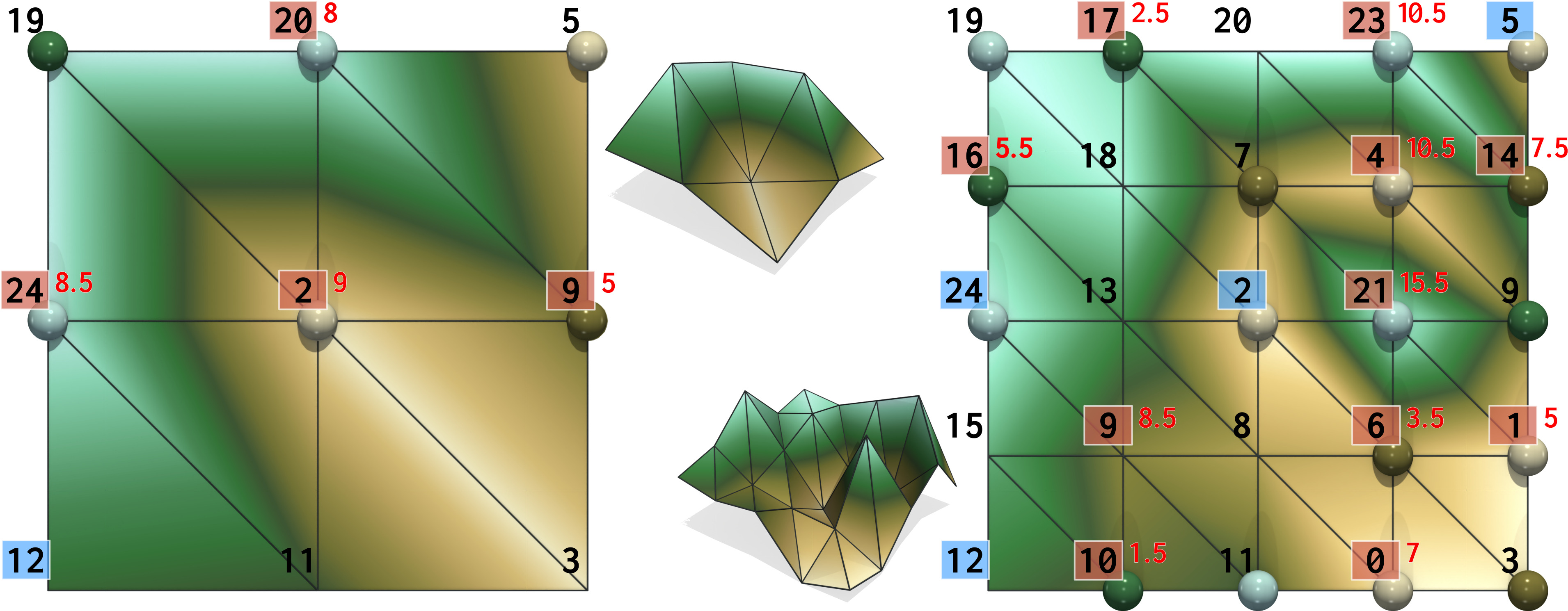}
        \hfill
        \includegraphics[height=0.18\linewidth]{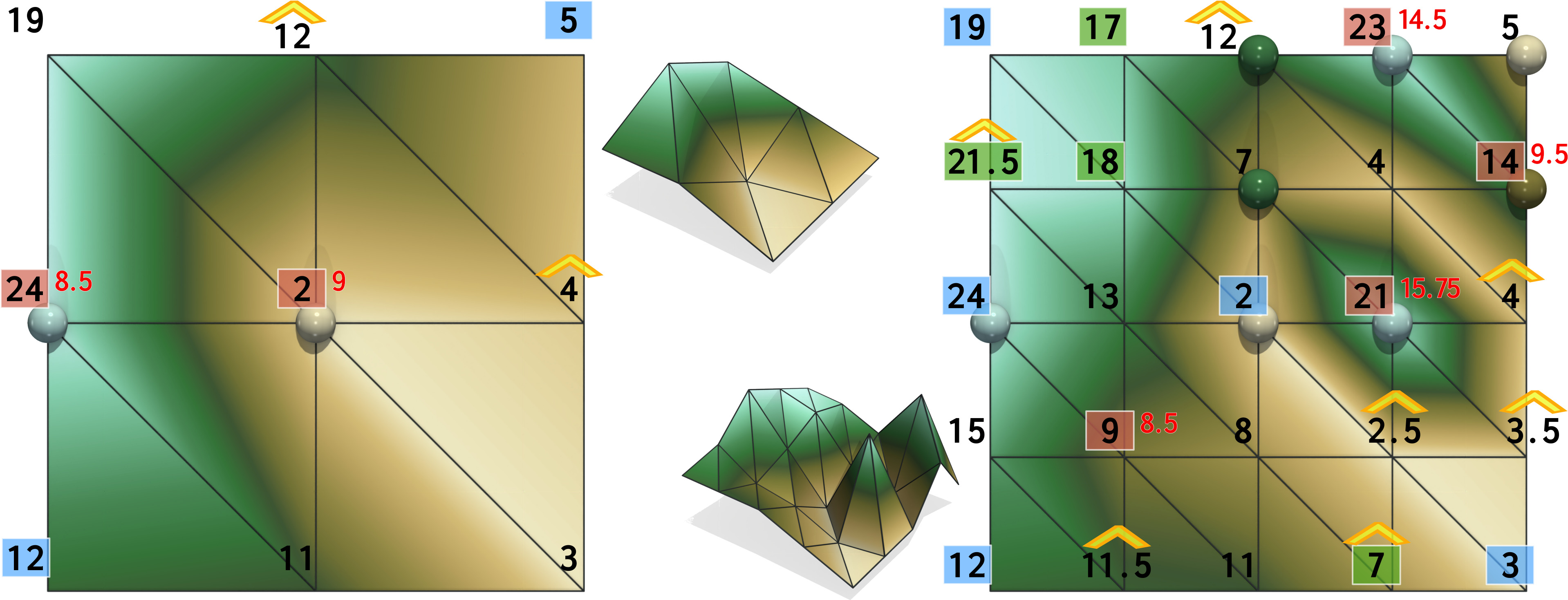}
        \mycaption{Traversal of the hierarchy and identification of topological
        invariants vertices for two different folding threshold (left:
        \julesEdit{$\epsilon=0\%$, right: $\epsilon=30\%$}).
The numbers denote the values of the approximated scalar field $\widehat{f}$.
Red squares indicate the non-monotonic new vertices, whose folding error
$\delta_{\epsilon}$ is labelled in red.
On the left, no vertex is folded and no approximation is made on the scalar
field (i.e. $\widehat{f}=f$). It results in a
high number of non-monotonic vertices and a low number of topologically
invariant old and new vertices (respectively blue and green squares).
In contrast, a folding threshold \julesEdit{$\epsilon=30\%$} is applied on the right. Every
non-monotonic vertex $n$ with a folding error
$\delta(n)\leq8$ gets folded (yellow hats). This reduces the number
of
non-monotonic vertices and more than doubles
the number of TI vertices (blue and green squares) of the full precision
approach.
}
\label{fig_vertex_folding}
\end{figure*}

\subsection{Vertex Folding}
\label{sec_vertex_folding}


This section describes the variations of the above strategy for the case where
$\epsilon\neq 0$, to decrease the computational cost.
Specifically, we present a
strategy to artificially
increase the number of topologically
invariant vertices
during the hierarchy traversal (step 1),
which will consequently result in skipping the estimation of vertex
criticality (step 2) for a larger number of vertices (hence the overall
speedup). 

A new vertex $n$, appearing at a
level $i$ of the hierarchy,
 is inserted at the center of
an old edge $(o_0,o_1)$ of $\domain^{i-1}$ (\autoref{sec_prelis_hierarchy}).
Assuming that $f(o_0)<f(o_1)<f(n)$, $n$ is a non-monotonic vertex and
impacts the link polarity of $o_1$.
The apparition of such vertices reduces the overall performance as these will
trigger explicit criticality computations in step 2.
To reduce
the number of non-monotonic vertices inserted in $\domain^i$,
we can choose to reinterpolate a non-monotonic  vertex $n$ between its two new
neighbors to enforce its monotony.
We note the resulting monotonic vertex $\widehat{n}$ and say that this vertex is
\textit{folded}. We define its new approximated value $\widehat{f}(\widehat{n})$
as
the interpolation of the approximated values of its two old neighbors:
$\widehat{f}(\widehat{n})=\left(\widehat{f}(o_0)+\widehat{f}(o_1)\right)/2$.
The values $\widehat{f}(o_0)$ and $\widehat{f}(o_1)$ are themselves either the result of a linear interpolation
if $o_0$ or $o_1$ have been previously \textit{folded}. Otherwise, they are
equal to $f(o_0)$ and $f(o_1)$.
\julesEdit{The linear interpolation is preferred here to alternatives as it provides a good balance between accuracy and efficiency.}

Formally, we build a sequence
$\{\widehat{f^0},\widehat{f^1},\hdots,\widehat{f^h}\}$ of PL scalar fields
defined
on each hierarchy level. The sequence is defined recursively:

\vspace{-1em}
\begin{enumerate}
        \itemsep0em
        \item{$\widehat{f^0}=f^0$}  
        \item{For each old vertex $o$ of $\domain^i$,
                $\quad\widehat{f}^i(o)=\widehat{f}{}^{i-1}(o)$}  
        \item{For each new vertex $n$ of $\domain^i$ that is not
                \textit{folded},\quad$\widehat{f^i}(n)=f^{i}(n)$}  
        \item{For each \textit{folded} new vertex $\widehat{n}$ of $\domain^i$
        on the edge $(o_0,o_1)$,
                $$\widehat{f^i}(\widehat{n})=\frac{\widehat{f^{i-1}}(o_0)+\widehat{f^{i-1}}(o_1)}{2}$$}
\end{enumerate}
\vspace{-1em}

We note $\foldedSet^i$ the set of folded vertices at level $i$ (with $\foldedSet^i \subset \domain_0^i$).
By construction, folded new vertices are monotonic.
\autoref{fig_vertex_folding} illustrates how a higher amount of folded vertices at level $i$ implies
a higher number of topologically invariant vertices identified on $\domain^i$.
In particular, if all non-monotonic vertices are folded at level $i$, all vertices of $\domain^i$ are topologically invariant.

%

\subsection{Bottleneck Error Control}
\label{sec_error_control}
Computing persistence diagrams
with \textit{vertex folding}
results in approximations of the exact result $\diagram(f)$,
given by $\diagram(\widehat{f})$ (diagram of the approximated
field $\widehat{f}$). Then the resulting approximation error (in terms of
Bottleneck distance) is given by \cite{CohenSteinerEH05}:
%
$W_\infty\big(\diagram(\widehat{f}),\diagram(f)\big)\leq\|\widehat{f}-
f\|_\infty$ , which is rather easy to estimate.
%
For each
new vertex $n$ inserted at level $i$ of the hierarchy on the edge $(o_0,o_1)$,
we define its \textit{folding error} $\delta(n)$ as the difference between its
original scalar value and its reinterpolation value at its level of insertion:
$\delta(n)=\left|\left(\widehat{f^i}(o_0)+\widehat{f^i}(o_1)\right)/2-f(n)\right
|$.
Then, we have:
\vspace{-1ex}
\[
        \|\widehat{f}-f\|_\infty=\max_{\widehat{n}\in\foldedSet^h} \delta(\widehat{n})=\max_{\widehat{n}\in\foldedSet^h} |\widehat{f}(\widehat{n})-f(\widehat{n})|
        \vspace{-1ex}
\]

In the light of these observations, we use the following folding strategy as we process the hierarchy. 
Given a target approximation error $\epsilon$,
%
the hierarchy is processed as described
in \autoref{sec_hierarchy_processing},
except that vertex polarity is estimated from the approximated field
$\widehat{f}$.
For each level $i$, we choose to fold non-monotonic vertices $n$ with an error $\delta(n)<\epsilon$. Monotonic vertices or
non-monotonic vertices with a higher folding error get added into the hierarchy
without being reinterpolated.
Let $\widehat{f}_\epsilon : \domain \rightarrow \range$ be the final field
approximation (after the hierarchy traversal is completed), given
the target approximation error $\epsilon$.
Then
it is clear
that $\forall
v\in\domain,~\delta(v)\leq\epsilon$. Thus:
\julesEdit{
\vspace{-1ex}
\[
W_\infty\big(\diagram(\widehat{f}_\epsilon),\diagram(f)\big)\leq\|
\widehat{f}_\epsilon - f\|_\infty = \max_{\widehat{n}\in\foldedSet^h} \delta(\widehat{n}) \leq \epsilon
\vspace{-1ex}
\]
The computation must reach the last level $h$, to accurately capture $\max \delta(\widehat{n})$ (required for guarantees), thus it is not progressive.}
In the remainder, $\epsilon$ is noted as a percentage of the data range
(\julesEdit{$\epsilon=0\%$} is the exact computation, \julesEdit{$\epsilon=100\%$}
folds all non-monotonic vertices).


\begin{figure*}
        \centering
        \includegraphics[width=\linewidth]{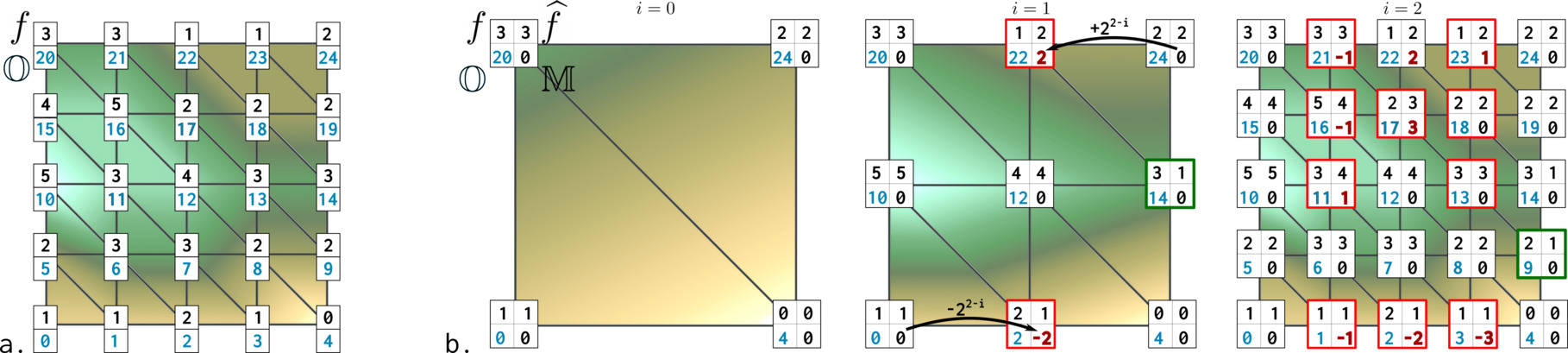}
                \mycaption{Monotony offsets on a toy example of an
\textbf{integer}
field
$f$ (a: black numbers) defined on a 2D grid.
                The injectivity of $f$ is guaranteed by the offset field
$\offsets$ (a: blue numbers).
                As the hierarchy is processed (b), some vertices get folded
                (red and green squares) according to a given error
threshold. Due to the precision
                of the field (here integer precision), their interpolated value
(numbers on the right) might be
                identical to that of an old neighbor.
                In some cases (red squares), the
                folding actually entails a monotony change as the offset field
                $\offsets$ (blue numbers)
                provides the wrong order relation between neighbor vertices. If
such an event occurs, the
                monotony offset of the folded vertex is updated (red numbers) to
                enforce monotony.
                Green squares denote folding cases where
$\offsets$ does not contradict the folding monotony.
                At the finest resolution (b, rightmost), this
                results
in plateaus (bottom row of the grid) where
                the injectivity of $\widehat{f}$ is guaranteed and where
                the monotony is correctly enforced.
                }
\label{fig_monotony_offsets}
\end{figure*}
\subsection{Monotony offsets}
\label{sec_monotonyOffsets}
We now discuss how to handle degenerate flat plateaus, which
become more frequent
given our vertex folding strategy.
The input scalar field $f$ is
injective
on the vertices of $\domain$ (\autoref{sec_preliminaries}). This is enforced in
practice with
an
offset field $\offsets: \domain_0 \rightarrow \mathbb{N}$,
typically corresponding for each vertex to its offset in memory
(\autoref{fig_monotony_offsets}a).
$\offsets$ is then used
to disambiguate vertices with identical scalar values.

In theory, a \textit{folded} vertex $\widehat{n}$ is guaranteed to be monotonic.
However in practice, if $\delta(\widehat{n})$ falls below the precision of the
data type used to encode the scalar field, a flat plateau emerges. This occurs
frequently for instance when the input data is expressed with integers.
%
Then, given a new folded vertex $\widehat{n}$ inserted on an edge
$(o_0, o_1)$, we may have:
%
$\widehat{f}(\widehat{n})=\widehat{f}(o_0)$, which means
$\widehat{n}$ and $o_0$ will be disambiguated in the algorithm
by their offset $\offsets$.
However,
this can introduce undesired monotony changes
(\autoref{fig_monotony_offsets}b, red squares).

To guarantee the monotony of folded vertices,
%
we introduce
a \textit{monotony offset} on each vertex $v$, noted $\monotonyOffset(v)$,  which is modified when the vertex gets folded.
The purpose of the monotony offset $\monotonyOffset$ is to take over the regular
offset $\offsets$ if it contradicts the monotony of newly folded vertices.
Given a new \textit{folded} vertex $\fold{n}$ inserted on an edge $(o_0,o_1)$ at level $i$, that is
non-monotonic with respect to $o_0$ (i.e.
$\widehat{f}(\fold{n})<\widehat{f}(o_0)<\widehat{f}(o_1)$ or
$\widehat{f}(\fold{n})>\widehat{f}(o_0)>\widehat{f}(o_1)$ ), we set:

\vspace{-3ex}
\[
\monotonyOffset(\fold{n})=
\left\{
\begin{aligned}
        &\quad\monotonyOffset(o_0) - 2^{h-i} \quad&&
        \small{
        \begin{aligned}
                \text{if} &\quad \widehat{f}(\fold{n})<\widehat{f}(o_0)<\widehat{f}(o_1)\\[-.5em]
                \text{and} &\quad \mathbb{O}(\fold{n})>\offsets(o_0)\\
\end{aligned}}\\
        &\quad\monotonyOffset(o_0) + 2^{h-i} \quad &&
        \small{
        \begin{aligned}
                \text{if} &\quad \widehat{f}(\fold{n})<\widehat{f}(o_0)<\widehat{f}(o_1)\\[-0.5em]
        \text{and} &\quad \mathbb{O}(\fold{n})<\offsets(o_0)\\
        \end{aligned} } \qquad\qquad\\
        &\quad\monotonyOffset(o_0) \quad && \small{\text{else}}\\
\end{aligned}
\right.
\vspace{-1ex}
\]

The monotony offset is initially set to zero for all new vertices and modified
only in
case of vertex folding.
Then,
the field $\monotonyOffset$ explicitly encodes
the monotony of newly
folded vertices (\autoref{fig_monotony_offsets}). The monotony
offsets
are used to disambiguate the comparison of two vertices of identical
approximated scalar value in the rest of the approach (criticality estimation,
integral lines, etc).

\subsection{Parallelism}
\label{sec_parallelism}

Our approach can be easily parallelized using shared-memory parallelism.
The first step of our approach, the traversal of the hierarchy, can be trivially parallelized on the vertices, 
as all operations are local to a vertex
\julesEdit{(\autoref{sec_vertex_folding})}. However the hierarchy must be
processed in sequential, which implies synchronization 
between the different levels. The computation of the criticality for non topologically invariant vertices is also completely parallel.
The computation of the persistence diagram from the critical points can be
parallelized on the saddle points, however locks are necessary for the parallel
computation of integral lines, in order to back-propagate the indices of
found extrema. The saddle points are sorted in parallel, using the GNU implementation \cite{singler2008gnu}.
Finally, processing the merge events represented by the triplets is a sequential task, but represents only a fraction
of the total sequential computation time (less that~1\%~in~practice).

\subsection{Uncertainty}
\label{sec_uncertainty}

\begin{figure}
        \includegraphics[width=\linewidth]{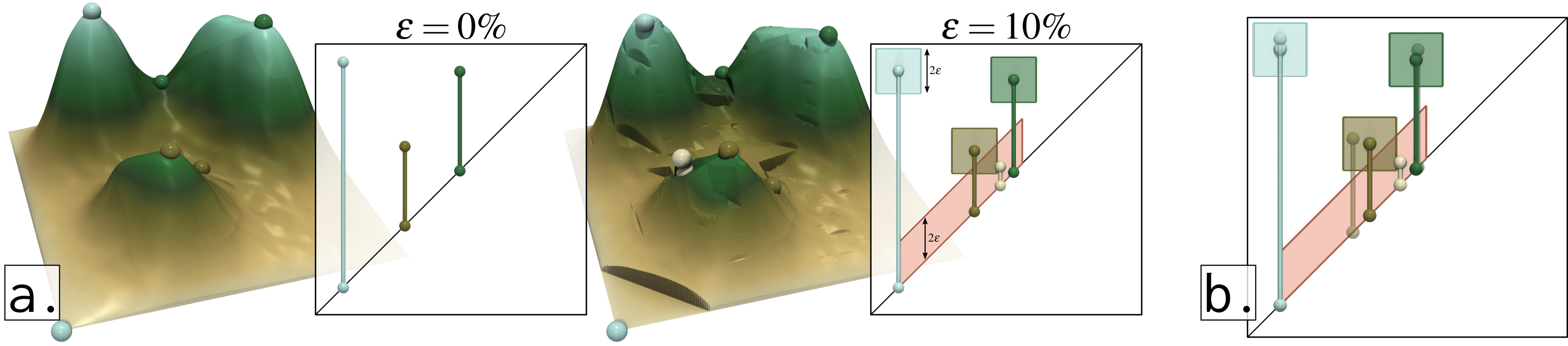}
        \mycaption{\julesEdit{
        Approximation uncertainty visualization.
        The 
        diagram at
        $\epsilon=10\%$ (a., right) exhibits
        one \textit{uncertain} pair within the red band, which is absent from the exact result (a., left). Squares bound the correct location of \textit{certain} pairs (b., transparent: exact pairs).}}
        \label{fig_uncertainty_visu}
        \vspace{1ex}
\end{figure}

By construction, our approach induces a Bottleneck error of $\epsilon$, which
corresponds to a maximum mismatch of $\epsilon$ between the pairs
of $\diagram(\widehat{f}_\epsilon)$ and these of $\diagram(f)$.
This means that
some approximated persistence pairs
(with a persistence below $2\epsilon$)
may not be present in the exact diagram (as they may be matched to the
diagonal at a cost lower than $\epsilon$).
We call these pairs \textit{uncertain}. In contrast,
the approximated pairs with
a persistence beyond $2\epsilon$ will certainly be present in the exact
diagram, and their exact location
is bounded within a square of side $2\epsilon$. We call these pairs
\textit{certain}.
Thus, we can visually convey the level of uncertainty of our approximations
directly in the persistence diagrams
\julesEdit{(\autoref{fig_uncertainty_visu}}). For this, we use a red band to
indicate
\textit{uncertain} pairs, and we draw the bounding squares
for \textit{certain} pairs (\autoref{sec_visu}).



\section{Results}
\label{sec_results}

This section presents experimental results obtained on a variety of datasets,
available on public repositories \cite{openSciVisDataSets, ttkData}. 
We implemented our approach in C++, as modules for the Topology ToolKit (TTK)
\cite{ttk17, ttk19}.
The experiments were carried out on a desktop computer with two Xeon CPUs (3.0 GHz,
2$\times$4 cores) and 64 GB of RAM.


\subsection{Time Performance}

\begin{table}
        \centering
        \caption{Increase of the number of topologically invariant (TI)
                vertices for different levels of approximation. For
                real-world datasets (\textit{MinMax} and \textit{Random}
excluded), the average proportion
                of TI vertices rises from 70\% for the full precision approach
                (\textit{i.e. \julesEdit{$\epsilon=0\%$}})
                to 94\% for a relative Bottleneck error of 5\% (\julesEdit{$\epsilon=5\%$}).}
        \rowcolors{3}{gray!20}{white}
        \resizebox{\columnwidth}{!}{
                \begin{tabular}{|l|r|rrrr|}
\hline
Dataset & $\sum_{i=0}^h |\domain^h_0|$  & \multicolumn{4}{c|}{\% TI}\\ &  &  \julesEdit{$\epsilon=0\%$} & \julesEdit{$\epsilon=1\%$} &\julesEdit{$\epsilon=5\%$} & \julesEdit{$\epsilon=10\%$} \\\hline
At & 931,102 & 64.7 & 95.3 & 96.1 & 96.3\\
SeaSurfaceHeight & 1,384,636 & 61.4 & 90.8 & 96.7 & 98.3\\
Ethanediol & 2,057,380 & 77.7 & 97.4 & 97.6 & 97.6\\
Hydrogen & 2,413,516 & 73.5 & 97.8 & 97.3 & 97.4\\
Isabel & 3,605,596 & 44.5 & 80.8 & 91.6 & 93.5\\
Combustion & 4,378,378 & 65.5 & 89.6 & 96.3 & 97.3\\
Boat & 4,821,318 & 88.5 & 96.9 & 97.0 & 97.2\\
MinMax & 18,994,891 & 99.3 & 99.5 & 99.5 & 99.5\\
Aneurism & 19,240,269 & 95.6 & 96.1 & 97.4 & 98.2\\
Foot & 19,240,269 & 64.4 & 66.6 & 73.7 & 86.9\\
Heptane & 31,580,914 & 82.2 & 96.4 & 98.2 & 98.7\\
Random & 18,117,510 & 0.9 & 0.9 & 1.0 & 1.1\\
Backpack & 111,929,613 & 39.1 & 77.2 & 94.7 & 97.7\\
\hline
\end{tabular}

        }
        \label{table_stats}
\end{table}
\begin{table}
        \caption{Sequential computation times (in seconds) of our approach for
the approximation
        of persistence diagram, for different approximation errors. The
column \emph{Default} reports the run time
        of the default approach in the Topology ToolKit \cite{gueunet_tpds19}.
The last column
        indicates the speedup of our approach with an approximation error of
$5\%$, against
        the fastest of the two exact methods (left).
\julesEdit{Bold numbers indicate the smallest $\epsilon$ providing a speedup over reference approaches.}}
\rowcolors{3}{gray!20}{white}
        \resizebox{\columnwidth}{!}{
        \begin{tabular}{|l|rr|rrr|r|}
\hline
Dataset & Default\cite{gueunet_tpds19} &Progressive\cite{vidal21} &
\multicolumn{4}{c|}{Ours} \\
 & & & $\epsilon=1\%$ & $\epsilon=5\%$ & $\epsilon=10\%$ & 5\% speedup\\
\hline
At & 0.27 & 0.25 & \textbf{0.14} & 0.15 & 0.17 & 38.5\%\\
SeaSurfaceHeight & 0.48 & 0.38 & \textbf{0.29} & 0.26 & 0.25 & 30.9\%\\
EthaneDiol & 0.48 & 0.43 & \textbf{0.28} & 0.31 & 0.33 & 28.7\%\\
Hydrogen & 0.99 & 0.64 & \textbf{0.43} & 0.48 & 0.47 & 24.6\%\\
Isabel & 1.29 & 1.49 & \textbf{0.95} & 0.89 & 0.92 & 30.7\%\\
Combustion & 2.55 & 1.37 & \textbf{0.99} & 0.90 & 0.85 & 34.1\%\\
Boat & 1.22 & 0.82 & 0.97 & 1.06 & 1.02 & -28.1\%\\
MinMax & 4.01 & 1.92 & 2.14 & 2.28 & 2.13 & -18.4\%\\
Aneurism & 4.66 & 3.43 & 3.62 & 3.52 & \textbf{3.00} & -2.7\%\\
Foot & 9.86 & 10.42 & 10.38 & \textbf{8.14} & 6.14 & 17.4\%\\
Heptane & 8.09 & 7.41 & \textbf{5.41} & 5.18 & 5.16 & 30.1\%\\
Random & 37.29 & 30.77 & \textbf{28.95} & 29.04 & 30.46 & 5.6\%\\
Backpack & 77.28 & 107.31 & \textbf{62.06} & 40.11 & 31.76 & 48.1\%\\
\hline
\end{tabular}

}
\label{table_perf_seq}
\vspace{-2ex}
\end{table}


The time complexity of our approach is similar to the complexity of the
non-interruped algorithm
by Vidal et al. \cite{vidal21}.
The first two steps of our approach, that amount to the hierarchy traversal and the computation of
critical points, have a linear complexity in the number of vertices:
$\mathcal{O}(\sum_{i = 0}^{i= h} |\domain_0^i|)$.
The third step, to compute the saddle-extremum persistence from the critical points, is identical to
the approach by Vidal, except that we compute the persistence
diagram exclusively on the last, finest level of the hierarchy.
It has a practical time complexity \cite{vidal21} of 
$\mathcal{O}\big( |\domain_1^h| + n_s\log n_s + n_s\alpha(n_m)\big)$, 
where $\alpha$ stands for the inverse of the Ackermann function, and $n_s$ and $n_m$ respectively denote the
number of saddle points and extrema.

Table \ref{table_stats} reports the number of TI vertices for all datasets,
for various
approximation errors.
The column $\epsilon=0$ corresponds to the
numbers of TI vertices reported by Vidal et al. \cite{vidal21} in the exact
case.
For the majority of datasets, we observe a large increase in the number of TI vertices, 
from a proportion of 70\% on average on the real-life datasets to a proportion of 90\% for
a small error of $1\%$, and 94\% for a mild tolerance of $5\%$ on the
approximation. The largest increase in the number of TI vertices is reported for
\textit{Backpack} (being a large and noisy dataset).
This table confirms that our strategy of \textit{vertex folding}
(\autoref{sec_vertex_folding})
indeed implies a sensible increase in the number of TI vertices, even for mild
approximation errors.
Regarding the criticality estimation (step 2,
\autoref{sec_hierarchy_processing}),
as
no computation is needed for the
vertices which
remained
topologically invariant throughout the hierarchy
(\autoref{sec_hierarchy_processing}), a higher proportion of TI
vertices in the data is thus likely to significantly
decrease the computational workload, resulting in lower computation times.

Table \ref{table_perf_seq} details the sequential computation times of our
approach for different approximation errors.
They are compared with public implementations of exact algorithms, both
available in TTK \cite{ttk17}:
the \emph{progressive} approach by Vidal et al.\cite{vidal21} (run up to the
finest resolution, hence producing an exact result), and the \emph{default}
algorithm used in TTK \cite{gueunet_tpds19} (run at the finest hierarchy level).
The last column of \autoref{table_perf_seq} present the speedups obtained with
a Bottleneck error tolerance of 5\%,
compared with the fastest of either reference approaches. We observe an average
reduction of the run times of 18\% on real-world datasets, which confirms that
our strategy of maximizing the number of TI vertices
effectively reduces the computation times.
The observed speedups are consistent with the increases in the proportion of
TI vertices reported in \autoref{table_stats}: a large increase in the number
of TI vertices implies an important reduction of the computation time.
Interestingly, we find our method most beneficial on datasets that initially present a low amount of TI vertices.
This usually corresponds to a high level of noise, which impedes both reference
methods. In particular, the highest speedup
is achieved on \textit{Backpack}, our largest, noisiest real-world dataset,
with a reduction of computation time of nearly 50\%.
In contrast, our method fails to reduce the computation times on smooth
datasets such as \textit{MinMax}, \textit{Boat} or \textit{Aneurism}, 
for which the \emph{progressive} approach
really shines.
\julesEdit{These datasets exhibit high initial proportions of TI vertices, which limits the
increase in TI vertices enabled by our approach (\autoref{table_stats}).}
For \textit{Random},
the interpolation cost of folded vertices
seems to counterbalance
the speedup induced by the increase in TI vertices.

Table \ref{table_perf_para} lists the computation times obtained with the
parallel version
of our algorithm.
We find an
overall
average parallel efficiency of 43\% with an error level of 5\%, which is on par
with the progressive approach  \cite{vidal21}.
Although the traversal of the hierarchy can be trivially parallelize over vertices, it is subject to 
synchronization steps between
hierarchy levels.
The last step of the approach,
deducing the persistence diagram from the critical points,
is less balanced. Indeed, the parallel computation of integral lines between saddle
points and extrema (\autoref{sec_hierarchy_processing}) necessitates locks.

\begin{table}
        \caption{Parallel computation times (in seconds, on 8 physical cores) of 
our algorithm with different approximation errors.
        The presented speedups relate to the sequential run times.}
\rowcolors{3}{gray!20}{white}
        \resizebox{\columnwidth}{!}{
        \begin{tabular}{|l|rr|rr|rr|rr|}
\hline
Dataset & Default\cite{gueunet_tpds19} & Progressive\cite{vidal21} &
\julesEdit{$\epsilon=1\%$} &speedup & \julesEdit{$\epsilon=5\%$} & speedup & \julesEdit{$\epsilon=10\%$} & speedup
\\
\hline
At & 0.20 & 0.07 & 0.05 & 2.61 & 0.07 & 2.24 & 0.07 & 2.25\\
SeaSurfaceHeight & 0.21 & 0.13 & 0.08 & 3.65 & 0.07 & 3.91 & 0.07 & 3.77\\
EthaneDiol & 0.21 & 0.11 & 0.08 & 3.69 & 0.09 & 3.60 & 0.10 & 3.35\\
Hydrogen & 0.72 & 0.20 & 0.13 & 3.18 & 0.16 & 2.95 & 0.17 & 2.80\\
Isabel & 0.47 & 0.46 & 0.54 & 1.77 & 0.28 & 3.15 & 0.30 & 3.05\\
Combustion & 0.34 & 0.51 & 0.54 & 1.83 & 0.31 & 2.92 & 0.25 & 3.41\\
Boat & 0.29 & 0.29 & 0.33 & 2.94 & 0.37 & 2.85 & 0.34 & 2.98\\
MinMax & 0.80 & 0.61 & 0.69 & 3.11 & 0.54 & 4.23 & 0.54 & 3.97\\
Aneurism & 2.03 & 1.51 & 1.62 & 2.23 & 1.36 & 2.59 & 1.33 & 2.25\\
Foot & 3.12 & 2.34 & 2.65 & 3.91 & 1.91 & 4.25 & 2.04 & 3.01\\
Heptane & 2.44 & 2.35 & 2.08 & 2.60 & 1.33 & 3.88 & 1.51 & 3.41\\
Random & 27.04 & 8.47 & 7.15 & 4.05 & 7.77 & 3.74 & 8.89 & 3.43\\
Backpack & 30.36 & 24.88 & 12.63 & 4.92 & 8.50 & 4.72 & 7.02 & 4.53\\
\hline
\end{tabular}
 
}
\label{table_perf_para}
\end{table}

\begin{figure}
        \includegraphics[width=\linewidth]{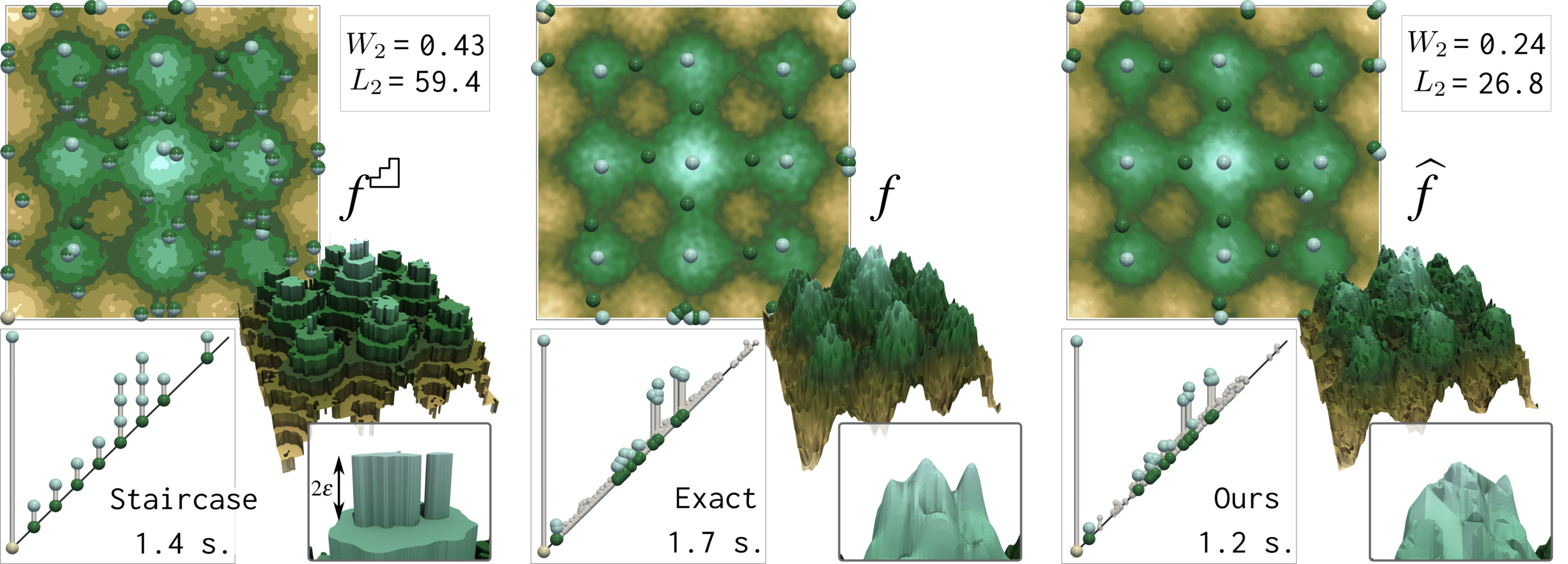}
        \mycaption{Comparison of our approach to a naive baseline
approximation (\textit{staircase} field),
                for the same tolerance of 5\% on the Bottleneck error.
Our approach (right)
provides an approximated persistence diagram that resembles the exact result
(center),
and is around 2 times closer in terms
of the Wasserstein distance ($W_2$).
The high persistence critical points (spheres, 2D domain) capture more
precisely
the features of the data in our case.
Our approximated field is also two times closer ($L_2$ norm) to the exact
field than the naive approximation.
}
\label{fig_staircase}
\end{figure}

\begin{table}[b]
        \vspace{-2.5ex}
        \caption{Accuracy comparison between our approach and a naive baseline
                approximation based on the computation of a \textit{staircase}
                function, for the same relative Bottleneck error of 5\%.
                Our approximations are more accurate in average, both in terms of the $L_2$
                distance of the approximated field (2 times more accurate)
                and in terms of the $L_2$-Wasserstein distance to the exact persistence diagram
        (5 times).}
        \rowcolors{3}{gray!20}{white}
        \resizebox{\columnwidth}{!}{
                \begin{tabular}{|l|rrr|rrr|}
\hline
Dataset & Staircase $L_2$ & Ours $L_2$ & Ratio & Staircase $W_2$ & Ours $W_2$ & Ratio\\
\hline
At & 276.66 & 75.31 & 3.67 & 3.31 & 1.01 & 3.29\\
SeaSurfaceHeight & 92.3 & 58.05 & 1.59 & 8.75 & 1.69 & 5.17\\
EthaneDiol & 337.55 & 73.2 & 4.61 & 1.78 & 0.61 & 2.9\\
Hydrogen & 11.0 & 13.0 & 0.85 & 25.69 & 12.73 & 2.02\\
Isabel & 3,591.99 & 1,569.98 & 2.29 & 42.26 & 8.08 & 5.23\\
Combustion & 38.04 & 17.59 & 2.16 & 0.87 & 0.21 & 4.1\\
Boat & 24.18 & 9.11 & 2.66 & 2.14 & 0.49 & 4.37\\
MinMax & 117.55 & 0.37 & 314.77 & 0.0 & 0.0 & -\\
Aneurism & 6.0 & 9.0 & 0.67 & 1,198.14 & 128.61 & 9.32\\
Foot & 16,006.11 & 10,784.33 & 1.48 & 5,859.67 & 1,419.51 & 4.13\\
Heptane & 5.0 & 12.0 & 0.42 & 716.1 & 61.12 & 11.72\\
Random & 29,270.73 & 3,468.14 & 8.44 & 23,863.56 & 1,038.1 & 22.99\\
Backpack & 142.0 & 142.0 & 1.0 & 17,941.13 & 1,933.83 & 9.28\\
\hline
\end{tabular}
 
        }
        \label{table_staircase_comparison}
\end{table}

\begin{figure}[t]
        \includegraphics[width=\linewidth]{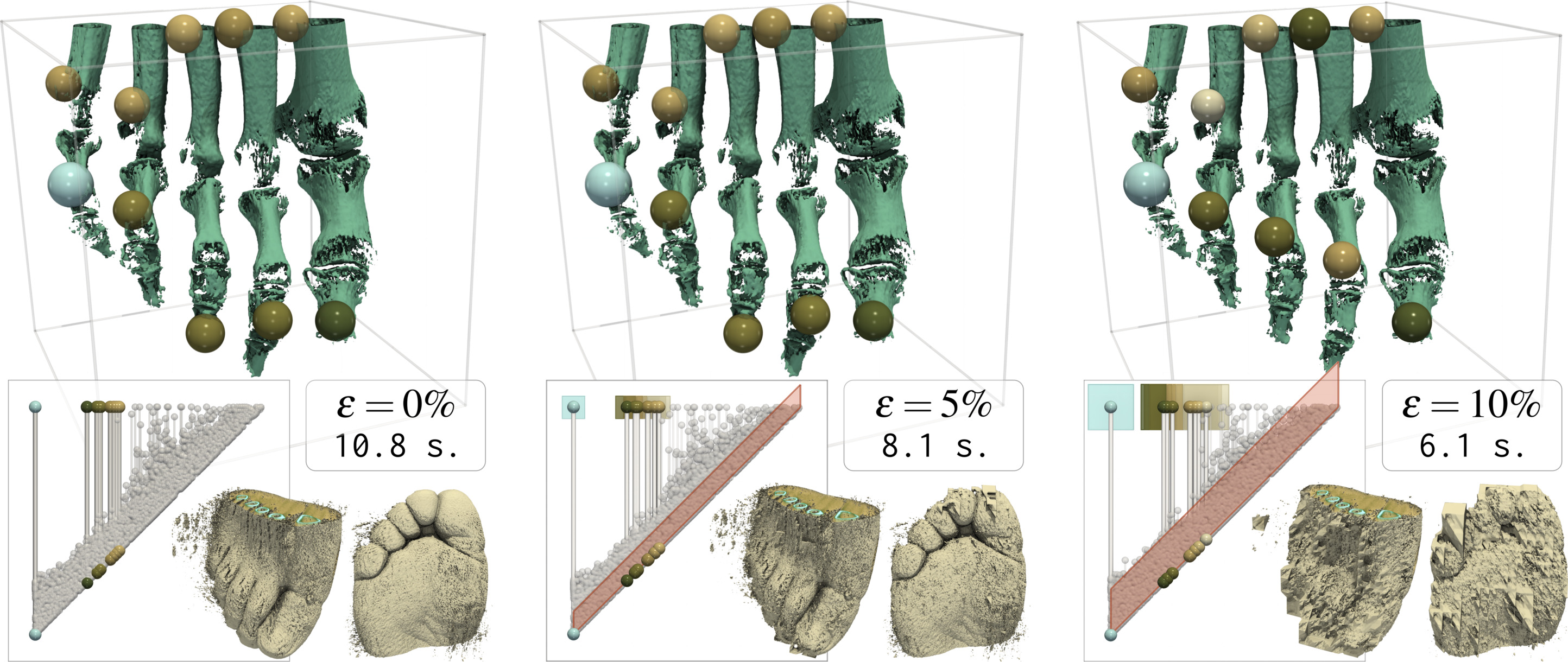}
        \mycaption{Approximated persistence diagrams on the \textit{Foot}
        dataset.
The 3D top views show the ten
most persistent maxima (spheres), corresponding to the bones of the foot
(isocontour, computed on the approximated field $\widehat{f}_\epsilon$).
Our approximated diagrams correctly capture the most persistent features
at a reduced computational cost.}
\label{fig_foot}
\vspace{-1ex}
\end{figure}

\begin{figure*}
        \includegraphics[width=\linewidth]{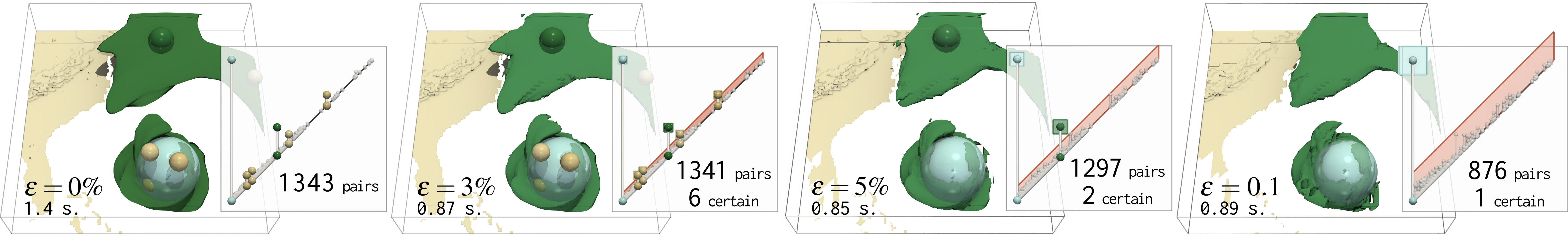}
        \mycaption{Approximated diagrams for the \textit{Isabel} hurricane
dataset. The most persistent
        maxima of the field (the magnitude of the wind velocity) are
represented
in the 3D view as spheres (scaled by persistence).
The approximation uncertainty is visualized in the diagram:
in
        the red band indicates \textit{uncertain} persistence
        pairs that may not exist in the exact persistence diagram, and colored
squares indicate
        a bound on the location of the persistence pairs that are
\textit{certain} to exist in the exact diagram.
        Our approximated persistence diagrams correctly capture these certain
pairs
        in the order of their persistence, with higher persistence features
        being detected at higher tolerance on the Bottleneck error.
}
\label{fig_isabel}
\end{figure*}

\subsection{Approximation Accuracy}

By design, our approach produces approximated persistence diagrams with a
guaranteed bound on the Bottleneck distance to the exact result
(\autoref{sec_error_control}).
In the following, we additionally evaluate the accuracy of our approximation by
considering
%
the $L_2$-Wasserstein distance (\autoref{sec_persistence_diagrams}, computed
with the efficient progressive approximation \cite{vidal20}
of TTK \cite{ttk17}) between our approximations and the
exact result.
We also evaluate $\|\widehat{f}-f\|_2$
to quantify the pointwise error of the approximated field $\widehat{f}$.
Results are given in Table \ref{table_staircase_comparison}.

For comparison, we perform the same evaluation for a naive baseline
approximation which provides identical guarantees.
This baseline consists in computing, with an exact algorithm, the diagram of a
\emph{staircase} function $\stairfield$, i.e. a quantized version   of the
input data $f$, with a quantization step of $2\epsilon$.
By construction, the staircase function
verifies $\|\stairfield-f\|_\infty<\epsilon$ (\autoref{fig_staircase})
and its
persistence diagram
is then guaranteed not to exceed a Bottleneck error of
$\epsilon$ \cite{CohenSteinerEH05}.
Table \ref{table_staircase_comparison}
compares the accuracy of this baseline approximation to our algorithm.
%
In terms of $L_2$ distance, our approximations of the input scalar fields
are around 2 times more accurate on average (on real-world datasets,
\textit{Random} and \textit{MinMax} excluded). This difference could be
expected,
as our method performs local and adaptive linear interpolations,
while the
staircase approach systematically flattens the data.
However, our approach is also significantly more accurate with
regards to the $L_2$-Wasserstein distance to the exact persistence diagrams: 4
times
in average on our real-world datasets.

Figure \ref{fig_staircase} further illustrates
the limitations of the \textit{staircase} baseline.
For a given approximation error
(5\%), our method gives an approximated
diagram
that is visually more similar to the exact result, and
which better depicts the number of salient features, as well as the noise in
the data.
%
In contrast, the diagram produced by the \textit{staircase}
approximation is more difficult to interpret, as the positions of persistence
pairs are quantized on a grid in the 2D birth/death  space, resulting in several
co-located pairs, which cannot be distinguished visually.
%
Our approximation of the scalar field is
also closer to the exact field, both visually and
in terms of the $L_2$ norm, enabling more accurate
critical point approximations in the domain (top).

\subsection{Qualitative Analysis}
\label{sec_visu}
This section discusses the utility of our approximations from a qualitative point
of view, for data analysis and visualization purposes.
\autoref{fig:teaser} shows the result of our approach on the \textit{Backpack}
dataset. In this
example, our approximated diagrams correctly capture the high persistence maxima
of the scalar field, which correspond
to high density objects inside the bag (bottles, wires, and metallic parts of
the bag). The approximate field $\widehat{f}_\epsilon$
resulting from the vertex folding (\autoref{sec_vertex_folding}) is
more
precise in the vicinity of features of high persistence. Indeed, the volume rendering in the top views of \autoref{fig:teaser}
shows clearly the objects inside the bag (high persistence maxima),
while isocontours capturing the
cloth of the bag (\autoref{fig:teaser}, bottom) illustrate the deterioration of the field in this region.
The same phenomenon can be noted for the \textit{Foot} dataset (\autoref{fig_foot}). More vertices are folded in the 
vicinity of low persistence features, typically the skin of the foot, and
the level of deterioration of the field increases
with the approximation error.
Conversely,
high persistence features (bones of the toes) are well
captured.

The above observation
suggests that our method provides a better approximation for persistent
features, and a more degraded evaluation for less persistent structures, which
was an original motivation for our approach (to focus the computational
efforts on relevant structures).
%
This is confirmed in \autoref{fig:teaser}  by the number of
\emph{noisy}
pairs (of low persistence, within the red band) in the approximated diagrams,
which
is significantly lower than the amount
of
noisy
pairs (of identical persistence) in the exact diagram (\autoref{fig:teaser},
in parenthesis).

\begin{figure}[h]
        \includegraphics[width=\linewidth]{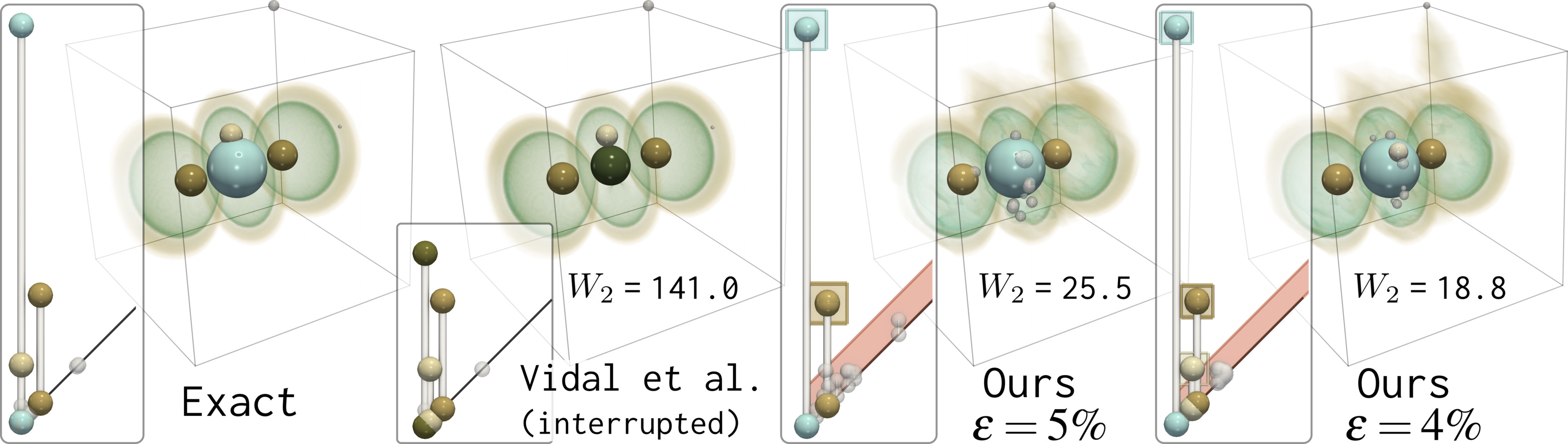}
        \mycaption{
        Approximation of persistence diagrams on the
\textit{hydrogen} dataset.
The best interrupted result of Vidal et al. \cite{vidal21}
                fails at correctly capturing the global
                maximum \julesEdit{(accurately detected only at the last level)}, resulting
                in a diagram that is 5 times less accurate than our 5\%
approximation ($L_2$-Wasserstein distance to the exact result). In
contrast, our approximation
                correctly captures the high persistence features of the data. On
the far right,
                our 4\% approximated diagram detects as \textit{certain} the
fourth most persistent maxima, which was
        marked as uncertain with \julesEdit{$\epsilon=5\%$}.
        }
        \label{fig_hydrogen}
        \vspace{-2ex}
\end{figure}

Figure \ref{fig_hydrogen}
compares our approximations to the progressive approach by Vidal et
al.\cite{vidal21}.
To generate an approximated result with the
progressive approach, we interrupt its computation at the penultimate hierarchy
level
(the computation would become exact at the final level).
As documented by its authors, such intermediate results
can provide useful previews, but however, with no guarantee
on the approximation error.
%
This is
illustrated in
\autoref{fig_hydrogen}, where the
 progressive approximation
fails at  correctly capturing the
maximum of largest persistence.
In contrast, our method produces persistence diagrams that correctly convey the
salience of the features, and are five times
more accurate, in terms of the Wasserstein distance.

Indications about the approximation uncertainty (\autoref{sec_uncertainty}) can
be displayed in the output diagrams.
Figure \ref{fig_isabel} shows our approximations for the \textit{Isabel}
dataset. For each approximation error, the red band indicates
uncertain pairs, which may not be part of the exact result.
\textit{Certain} pairs are represented with a square
bounding
their correct location. 
These
glyphs give a good sense of the approximation
uncertainty,
and are useful to assess the reliability of the diagram.
For instance, a
large pair in the uncertain zone
may indicate the presence of a medium persistence feature in the data.
This can be confirmed with 
the computation of a slighlty better estimation, as illustrated in
\autoref{fig_hydrogen} for the fourth most persistent feature.

\section{Limitations and Discussion}
\label{sec_limitations}
Our approximations tend to generate much less low-persistence features than
exact algorithms (\autoref{fig:teaser}), which can be an issue if features of
interested are hidden among noisy features near the diagonal.
On
the upside, this characteristic of our approximations make them well suited for
subsequent analysis and processing (e.g. distances and clustering), as diagrams
are often thresholded  in practice prior to further computations,
to remove low persistence pairs anyway.


An important limitation
of our approach, compared to the work of Vidal et al. \cite{vidal21}, is
its lack of progressivity.
Indeed, to provide strong approximation guarantees, the hierarchy has to be
completely traversed in our work and no intermediate result can be provided.
%

Another limitation is that our approach only supports saddle-extremum
persistence pairs at the moment. However, from our experience, these correspond
 in practice to the key features users tend to be interested in.
%

Finally, our approach provides strong guarantees on the Bottleneck distance.
Future work is needed for the theoretical study of
the impact of
our approximations on the $L_2$
Wasserstein
metric.

\section{Conclusion}
\label{sec_conclusion}
This paper introduced a method for the approximation of the persistence diagram
of a scalar field. Our work revisits the progressive approach by Vidal et
al.\cite{vidal21}, that
generated  preview
diagrams upon interruption of a progressive framework. We addressed the main drawback of their
approach, namely the lack of guaranteed error bounds on the diagram estimations.
In contrast, we presented a novel algorithm that efficiently computes the
approximation of a persistence diagram
within a user controlled approximation error on the Bottleneck distance to the exact result.
We showed that the approximated persistence diagrams are relevant for
visualization and data analysis tasks, as they correctly
describe the high persistence features in the data (\textit{i.e.} the number and salience of important features),
and they are more concise in practice than the exact diagrams. The uncertainty related to our approximations
can be effectively depicted visually inside the diagrams.

We believe that the development of approximative approaches (with guarantees)
for data
analysis and visualization
is an important and exciting research direction.
They are especially relevant in the field of Topological Data Analysis, 
as most of the computational workload is typically spent in practice on the
capture
of small scale
features (\autoref{fig:teaser}), whereas they usually are less relevant in
the applications, and post-process simplification techniques  are often applied
to eliminate them anyway.
In this context, a logical avenue for future work would be the development
of approximative
methods
to 
revisit existing topological data representations (merge trees, Morse-Smale
complexes, Reeb graphs).


\acknowledgments{
\small{
This work is partially supported by the
European Commission grant
ERC-2019-COG 
\emph{``TORI''} (ref. 863464, 
\url{https://erc-tori.github.io/}).
}
}


\bibliographystyle{abbrv-doi}

\clearpage

\bibliography{paper}
\end{document}